\documentclass[aps,showpacs,nofootinbib,prd,twocolumn,showkeys]{revtex4-1}
\usepackage{graphicx}
\usepackage{amssymb}
\usepackage{amsmath}
\usepackage[svgnames]{xcolor}
\usepackage{mathtools,slashed}
\usepackage[retainorgcmds]{IEEEtrantools}
\usepackage{physics}
\usepackage[compat=1.1.0]{tikz-feynhand}
\usepackage{tikz}
\usepackage{epstopdf}
\usepackage[utf8]{inputenc}
\usepackage{url}
\usepackage[colorlinks,citecolor=DarkGreen,linkcolor=DarkRed,urlcolor=DarkBlue]{hyperref}

\usepackage{newtxtext}
\usepackage{newtxmath}
\usepackage{hyperref}
\usepackage[makeroom]{cancel}
\usepackage{epsfig}

\newcommand{\be}{\begin{eqnarray}}
\newcommand{\ee}{\end{eqnarray}}
\begin{document}

\title{Two-gluon one-photon vertex in a magnetic field and its explicit one-loop  approximation in the intermediate field strength regime}

\author{Alejandro Ayala$^{1}$}
\author{Santiago Bernal-Langarica$^1$}
\author{Jorge Jaber-Urquiza$^2$}
\author{Jos\'e Jorge Medina-Serna$^1$}
  \address{
  $^1$Instituto de Ciencias
  Nucleares, Universidad Nacional Aut\'onoma de M\'exico, Apartado
  Postal 70-543, Ciudad de México 04510,
  Mexico.\\
  $^2$Facultad de Ciencias, Universidad Nacional Aut\'onoma de M\'exico, Apartado Postal 50-542, Ciudad de México 04510, Mexico.
  }
\begin{abstract}

We find the general structure for the two-gluon one-photon vertex in the presence of a constant magnetic field. We show that, when accounting for the symmetries satisfied by the strong and electromagnetic interactions under parity, charge conjugation and gluon interchange, and for gluons and photons on mass-shell, there exist only three possible tensor structures that span the vertex. These correspond to external products of the polarization vectors for each of the particles in the vertex. We also explicitly compute the one-loop approximation to this vertex in the intermediate field strength regime, which is the most appropriate one to describe possible effects of the presence of a magnetic field to enhance photon emission during pre-equilibrium in peripheral relativistic heavy-ion collisions. We show that the most favored direction for the photon to propagate is in the plane transverse to the field, which is consistent with a positive contribution to $\nu_2$ and may help to understand the larger than expected elliptic flow coefficient measured in this kind of reactions. 

\end{abstract}

\maketitle

\section{Introduction}\label{I}

The properties of strongly interacting matter in the presence of magnetic fields have become an important tool aiming to further our understanding of chiral symmetry and confinement dynamics. Theoretically, it has been found that short-lived but large intensity magnetic fields can be produced in peripheral heavy-ion collisions at high energies~\cite{Skokov:2009qp,Voronyuk:2011jd,McLerran:2013hla,Bzdak:2011yy,Zhang:2023ppo,Sun:2023rhh}. Experimentally, a peak intensity of order $B\sim 10^{19}$ G has been inferred at RHIC energies~\cite{STAR:2019wlg,Brandenburg:2021lnj}. The strength of these fields is usually taken as uniform, although it can in principle also be treated as white noise around a peak strength~\cite{Castano-Yepes:2023brq,Castano-Yepes:2022luw,Castano-Yepes:2024ctr}. In spite of the likely incomplete response of the medium produced by the fast decrease of the field intensity~\cite{Wang:2021oqq,Shovkovy:2022bnd}, its possible existence has prompted efforts to find signals of its presence and consequences in this environment. An important example of these efforts is the attempt to identify the charge separation signal along the magnetic field direction~\cite{STAR:2022ahj,Zhao:2022grq,STAR:2014uiw} with the chiral magnetic effect~\cite{Kharzeev:2007jp}. 

Another venue that has been explored is the possibility to link part of the excess yield and of the strength of the elliptic flow coefficient $\nu_2$ of direct photons
with effects associated to magnetic field induced processes. The large magnitude of  photon $\nu_2$, which is similar to that of hadrons\textcolor{black}{~\cite{PHENIX:2011oxq,ALICE:2018mjj,PHENIX:2015igl,ALICE:2018dti,David:2019wpt}}, is an intriguing property of the photon spectra in heavy-ion collisions which is referred to as the {\it photon puzzle}. Since hadron flow comes mainly from the late stages of the collision, the strength of the photon $\nu_2$ could be thought of as being produced also during the hadron part of the system's evolution. However, it has also been observed that for low transverse momentum ($p_T$) the photon yields show a thermal component that dominates over the prompt one that is even used to characterise the system's large temperature\textcolor{black}{~\cite{PHENIX:2014nkk,STAR:2016use,ALICE:2015xmh,PHENIX:2018che}} that, as such, should originate from the very early thermal history of the collision. An early emission of direct photons seems to be confirmed by the $p_T$ dependence of $\nu_2$ which, for large $p_T$ tends to zero. Indeed, when considering that photons, which are a penetrating probe, can only be boosted at the times when they are produced, then they should come from the early stages, where expansion velocities are small\textcolor{black}{~\cite{David:2019wpt}}. 

Measurements of the low $p_T$ excess of photons in Au+Au\textcolor{black}{~\cite{PHENIX:2014nkk,STAR:2016use}} and Pb+Pb\textcolor{black}{~\cite{ALICE:2015xmh}} over scaled p+p collisions at the same energy, have been complemented by recent PHENIX measurements in the smaller Cu+Cu system\textcolor{black}{~\cite{PHENIX:2018che}}, showing that the yield of these photons
scales with the number of binary collisions in Au+Au and Cu+Cu, which suggests that the source of these photons is similar for different colliding species and beam energies. The tension between the overall yields measured by STAR and PHENIX still remains. However, state of the art calculations for direct photon emission~\cite{Gale:2021emg} provide a good description of STAR and ALICE yields. Nevertheless, it should be pointed out that currently there is no approach that incorporates realistic dynamics obeyed by known sources of photons that can simultaneously reproduce the photon spectrum and elliptic flow. Future photon measurements in the lower NICA~\cite{MPD:2022qhn} energy domain promise to help clarify the picture.

Magnetic fields provide a source of electromagnetic radiation that at the same time induces a natural anisotropic emission and thus contributes to $\nu_2$ without the need to link its strength to the flow properties of the system. Some possible channels allowed by the presence of magnetic fields have been recently explored using a wide range of approaches. The emission of photons by magnetic field induced bremsstrahlung and pair annihilation in a quark-gluon plasma has been studied in Refs.~\cite{Tuchin:2014pka,Zakharov:2016mmc,Wang:2022jxx,Wang:2020dsr} and, accounting also for the effects of a slow rotation, in Ref.~\cite{Buzzegoli:2023vne}. Dilepton production in a magnetized QGP and a magnetized hadronic medium has been studied in Refs.~\cite{Wang:2021eud,Sadooghi:2016jyf} and~\cite{Mondal:2023vzx,Mondal:2023ypq}, respectively. Production of electromagnetic radiation from the QED$\times$QCD conformal anomaly has been studied in Ref.~\cite{Basar:2012bp}, from fluctuations of the gluon field coupled to the photon stress tensor in Ref.~\cite{Basar:2014swa}, from Cherenkov emission in a strong magnetic field in Ref.~\cite{Lee:2020tay} and from a quark-gluon plasma in a weak magnetic field coupled to the longitudinal dynamics in the background medium in Ref.~\cite{Sun:2023pil}. Holographic methods have also been used to describe photon production from a
strongly coupled plasma in the presence of intense magnetic
fields~\cite{Avila:2022cpa,Arciniega:2013dqa,Mamo:2012kqw,Wu:2013qja}. Signals of
deconfinement associated to photon production  in a long-lived chromomagnetic background, triggered in turn by a flash of a strong electromagnetic field, have been studied in Ref.~\cite{Nedelko:2022kjy}.

In a series of recent works we have explored the idea that the presence of a magnetic field during the pre-equilibrium stage of the collision opens the gluon fusion and splitting channels for photon production~\cite{Ayala:2017vex,Ayala:2019jey,Ayala:2022zhu}. Although the amplitudes for these processes are suppressed with respect to quark or anti-quark splitting and quark-anti-quark annihilation
amplitudes, which represent the leading perturbative matrix elements
for photon emission, gluon fusion and splitting are enhanced with respect to processes involving quarks since, at pre-equilibrium, the occupation number of quarks is suppressed with respect to gluons by a factor of $\alpha_s^2$~\cite{Baier:2000sb,Garcia-Montero:2019vju,Monnai:2019vup}. When accounting for Pauli blocking in the final state we see that processes involving quarks either in the initial or final state are overall suppressed making gluon fusion/splitting become an important channel for photon emission at pre-equilibrium. Other calculations aiming to find possible missing photon contributions during pre-equilibrium have also been recently discussed~\cite{McLerran:2014hza,Kurkela:2018vqr,Kurkela:2018wud,Kasmaei:2019ofu,Churchill:2020uvk,Garcia-Montero:2023lrd}. 

The approximations explored in Refs.~\cite{Ayala:2017vex,Ayala:2019jey} limit the accuracy of the calculation to the low $p_T$ part of the photon spectrum. On the other hand, the approximation explored in Ref.~\cite{Ayala:2022zhu} does not account for the full tensor structure with the symmetry properties to describe the matrix element for processes involving the two-gluon one-photon vertex in the presence of a magnetic field.The present work amends such shortcoming. We carry out a full-fledged analysis starting from first principles to find the correct tensor structure to describe the two-gluon one-photon vertex in the presence of a magnetic field. This vertex is an essential ingredient to compute the matrix element describing gluon fusion/splitting processes. We deduce the general tensor structure of this vertex for gluons and photons of arbitrary virtuality in terms of the possible polarization vectors and show that the structure drastically simplifies when restricting the calculation to on-shell gluons and photons. We use the deduced tensor basis to express the two-gluon one-photon vertex at one-loop order in the intermediate field strength regime, relevant to describe the photon yield and $\nu_2$ in the context of relativistic heavy-ion collisions during pre-equilibrium. A similar calculation of a vertex involving two massless vectors and a massive scalar in the presence of a magnetic field has been discussed in Ref.~\cite{Jaber-Urquiza:2023swn}. 

The work is organized as follows: In Sec.~\ref{II} we deduce the general structure of the two-gluon one-photon vertex in a constant magnetic field. We find the set of polarization vectors to express this tensor structure, using the properties of the strong and electromagnetic interactions under parity and charge conjugation and the symmetry between gluon exchange. We show that when the gluons and the photon are on mass-shell, the number of basis vectors becomes significantly smaller than when describing the case of arbitrary gluon and photon virtuality. In Sec.~\ref{III} we compute the explicit coefficients that multiply the tensor structure form the one-loop approximation to the two-gluon one-photon vertex for the case where the photon energy is larger than the magnetic field strength. In Sec.~\ref{IV} we analyze the behavior of the basis coefficients as functions of the gluon and photon energies as well as of the field strength and the angle between the direction of propagation of the photon and the direction of the magnetic field. We finally summarize and conclude in Sec.~\ref{concl}.

\section{General structure of the one-photon two-gluon vertex in a constant magnetic field}\label{II}

\begin{figure}[t]
\centering
\begin{tikzpicture}
\begin{feynhand}
\vertex (a) at (-1.5,1.5) {$a, \;\mu$}; \vertex  (b) at (-1.5,-1.5) {$b,\;\nu$};  \vertex[NEblob] (c) at (0,0) {B}; \vertex (d) at (2,0) {$\alpha$};
\propag [glu] (a) to [edge label' =$p_1$] (c);
\propag [glu] (b) to [edge label' =$p_2$] (c) ;
\propag [bos] (c) to [edge label =$q$] (d);
\end{feynhand}
\end{tikzpicture}
\caption{General representation of the two-gluon one-photon vertex. The shaded blob represents the effect of a magnetic field.}
\label{fig:vertex}
\end{figure}
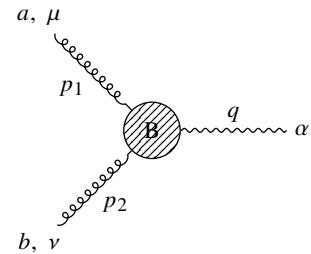
The two-gluon one-photon vertex in the presence of a magnetic field, depicted in Fig.~\ref{fig:vertex}, is denoted as
\begin{eqnarray}
\Gamma^{\mu\nu\alpha}_{ab}(p_1,p_2,q),
\label{genstructvert}
\end{eqnarray}
where $p_1,p_2$ are the gluon momenta and $q$ the photon momentum, $\mu\ ,\nu$ and $\alpha$ the Lorentz indices for the gluons and photon, respectively and $a,b$ the color indices of the two gluons. In what follows, we start from first
principles and concentrate in obtaining the tensor structure for the general gluon and photon off-mass shell case, to then specialize on determining the tensor structure for the on-mass shell case, which, as we will see, becomes a simpler exercise when starting from the general case. Notice also that in the absence of the magnetic field, the vertex vanishes, as demanded by Furry's theorem.

The vertex is a third rank tensor which is constrained by symmetry properties. First, gauge invariance requires that this tensor be transverse when contracted with the momentum carrying the Lorentz index corresponding to the vertex index assigned to the given particle, namely,
\begin{eqnarray}
p_{1\mu}\Gamma^{\mu\nu\alpha}_{ab}(p_1,p_2,q)&=&0\nonumber\\
p_{2\nu}\Gamma^{\mu\nu\alpha}_{ab}(p_1,p_2,q)&=&0\nonumber\\
q_{\alpha}\Gamma^{\mu\nu\alpha}_{ab}(p_1,p_2,q)&=&0.
\label{transversality}
\end{eqnarray}
Second, since the gluons are indistinguishable, the vertex needs to be symmetric under the exchange of the gluon indexes and momenta, namely,
\begin{eqnarray}
\Gamma^{\mu\nu\alpha}_{ab}(p_1,p_2,q)=\Gamma^{\nu\mu\alpha}_{ba}(p_2,p_1,q).
\label{exchange}
\end{eqnarray}
Third, since charge ($C$) and parity ($P$) conjugation are good quantum numbers for the strong and electromagnetic interactions, and three neutral gauge bosons are involved, we require that
\begin{eqnarray}
\hat{C}\ \Gamma^{\mu\nu\alpha}_{ab}\hat{C}^{-1}&=&(-1)^3\Gamma^{\mu\nu\alpha}_{ab}\nonumber\\
\hat{P}\ \Gamma^{\mu\nu\alpha}_{ab}\hat{P}^{-1}&=&(-1)^3\Gamma^{\mu\nu\alpha}_{ab}.
\label{CandP}
\end{eqnarray}
Notice that  Eqs.~(\ref{CandP}) imply that the vertex is invariant under the combined action of the $CP$ transformation.

The general properties in Eqs.~(\ref{transversality}),~(\ref{exchange}) and~(\ref{CandP}) need to be translated into the properties of the chosen tensor basis and its coefficients to span the vertex. The basis can be constructed out of the external product of suitably chosen polarization vectors. To find the tensor basis, let us first concentrate in the photon leg. We introduce a {\it tetrad}, namely, a set of four linearly independent vectors that span the four-dimensional Minkowski space-time and that also account for the presence of the external field. We start by choosing one of this vectors as the photon momentum and the other three as the general polarization vectors. We choose the Ritus basis~\cite{Papanyan:1971cvCORR,Papanyan:1973xaCORR} given by
\begin{eqnarray}
q^\mu&&\nonumber\\
l_q^\mu&\equiv&\hat{F}^{\mu\beta}q_\beta\nonumber\\
l_q^{*\mu}&\equiv&\hat{F}^{*\mu\beta}q_\beta\nonumber\\
k_q^\mu&\equiv&\frac{q^2}{l_q^2}\hat{F}^{\mu\beta}\hat{F}_{\beta\sigma}q^\sigma + q^\mu,
\label{Ritusbasis}
\end{eqnarray}
where 
\begin{eqnarray}
\hat{F}^{\mu\beta}\equiv F^{\mu\beta}/|B|,
\label{EMtensor}
\end{eqnarray}
with $F^{\mu\beta}$ the electromagnetic field strength tensor, $F^{*\mu\beta}$ its dual and $|B|$ the strength of the magnetic field. Notice that the polarization vectors $l_q^\mu,\ l_q^{*\mu}$ and $k_q^\mu$ satisfy
\begin{eqnarray}
q_\mu l_q^\mu=q_\mu l_q^{*\mu}=q_\mu k_q^\mu=0,
\label{poltransv}
\end{eqnarray}
and therefore are transverse to the photon momentum $q^\mu$.
The set of vectors in Eq.~(\ref{Ritusbasis}) form an orthogonal basis and satisfy the closure relation
\begin{eqnarray}
g^{\mu\nu}=\frac{q^\mu q^\nu}{q^2} + \frac{l_q^\mu l_q^\nu}{l_q^2} + \frac{l_q^{*\mu} l_q^{*\nu}}{l_q^{*2}} + \frac{k_q^\mu k_q^\nu}{k_q^2}.
\label{closure}
\end{eqnarray}
Notice that for on-shell photons $q^2=0$, and thus the vector $k^\mu$ becomes
\begin{equation*}
k_q^\mu\to q^\mu,
\end{equation*}
showing that the only two polarization vectors are $l_q^\mu$ and $l_q^{*\mu}$, which is consistent with the fact that for real photons there are only two polarizations. Also notice that in case the external field is taken in the $\hat{z}$-direction, the explicit form of the Ritus basis is given by
\begin{eqnarray}
q^\mu&=&(\omega_q,q_x,q_y,q_z)\nonumber\\
l_q^\mu&=&(0,q_y,-q_x,0)\nonumber\\
l_q^{*\mu}&=&(q_z,0,0,\omega_q)\nonumber\\
k_q^\mu&=&(\omega_q, (1-q^2/q_\perp^2)q_x, (1-q^2/q_\perp^2)q_y,q_z),
\label{explicitRitus}
\end{eqnarray}
where we define
\begin{eqnarray*}
q_\perp^2=-(q_x^2+q_y^2).
\label{qperpdef}
\end{eqnarray*}
Normalizing the polarization vectors in Eq.~(\ref{explicitRitus}) we obtain
\begin{eqnarray}
\hat{l}_q^\mu&=&\frac{1}{\sqrt{-q_\perp^2}}(0,q_y,-q_x,0)\nonumber\\
\hat{l}_q^{*\mu}&=&\frac{1}{\sqrt{q_\parallel^2}}(q_z,0,0,\omega_q)\nonumber\\
\hat{k}_q^\mu&=&\frac{1}{\sqrt{q^2q_\perp^2q_\parallel^2}}(q_\perp^2\omega_q, -q_\parallel^2q_x,-q_\parallel^2q_y,q_z),
\label{explicitRitusnorm}
\end{eqnarray}
where we define
\begin{eqnarray*}
q_\parallel^2=(\omega_q^2-q_z^2).
\label{qparadef}
\end{eqnarray*}
The polarization vectors in Eq.~(\ref{explicitRitusnorm}) coincide with the corresponding ones discussed in Ref.~\cite{Hattori:2017xoo}.

We now proceed to express the vertex in terms of the external product of the polarization vectors corresponding to each of the vector particles, namely, the two gluons and the photon. For this purpose, we can repeat the above process for each gauge particle to obtain tetrads associated to their momenta and thus their corresponding polarization vectors, namely
\begin{eqnarray}
{\mbox{photon}}\ \alpha&\to& q^\alpha,\ l_q^\alpha,\ l_q^{*\alpha},\ k_q^\alpha\nonumber\\
{\mbox{gluon}}\ \mu,\ a&\to& p_{1a}^\mu,\ l_{p_1a}^\mu,\ l_{p_1a}^{*\mu},\ k_{p_1a}^\mu\nonumber\\
{\mbox{gluon}}\ \nu,\ b&\to& p_{2b}^\nu,\ l_{p_2b}^\nu,\ l_{p_2b}^{*\nu},\ k_{p_2b}^\nu .
\label{forall}
\end{eqnarray}
Therefore, in general the $i$-th basis element
\begin{eqnarray*}
\Gamma^{\mu\nu\alpha}_{ab\ i}(p_1,p_2,q),
\end{eqnarray*}
corresponds to one of the products of three polarization vectors, where each factor is taken from one of the sets of polarization vectors for each particle. Working with the set of normalized polarization vectors, schematically we have
\begin{eqnarray}
\Gamma^{\mu\nu\alpha}_{ab\ i}(p_1,p_2,q)\!\!&\in&\!\! \left\{
\left[\hat{l}_{p_1a}^\mu,\ \hat{l}_{p_1a}^{*\mu},\ \hat{k}_{p_1a}^\mu
\right]
\otimes
\left[\hat{l}_{p_2b}^\nu,\ \hat{l}_{p_2b}^{*\nu},\ \hat{k}_{p_2b}^\nu
\right]\right.\nonumber\\
\!\!&\otimes&\!\!\left.
\left[\hat{l}_q^\alpha,\ \hat{l}_q^{*\alpha},\ \hat{k}_q^\alpha
\right]
\right\}.
\label{schematical}
\end{eqnarray}
In principle there are 27 possible basis vectors. We now proceed to use the restrictions imposed by Eqs.~(\ref{transversality}),~(\ref{exchange}) and~(\ref{CandP}) to reduce the number of basis elements.

Given the property expressed in Eq.~(\ref{poltransv}), we notice that Eq.~(\ref{transversality}) is automatically satisfied by each of the above $i=1,\ldots 27$ tensor structures in Eq.~(\ref{schematical}). The condition expressed in Eq.~(\ref{exchange}) is satisfied by symmetrizing the products of basis vectors in Eq.~(\ref{schematical}) under gluon exchange to obtain
\begin{eqnarray}
\Gamma^{\mu\nu\alpha}_{ab\ i}(p_1,p_2,q)&\in& \left\{
\left[\hat{l}_{p_1a}^\mu \hat{l}_{p_2b}^\nu , \hat{l}_{p_1a}^{*\mu} \hat{l}_{p_2b}^{*\nu},\hat{k}_{p_1a}^\mu \hat{k}_{p_2b}^\nu ,\right.\right.\nonumber\\
&&\frac{1}{\sqrt{2}}\left(\hat{l}_{p_1a}^\mu \hat{l}_{p_2b}^{*\nu} + \hat{l}_{p_1a}^{*\mu}\hat{l}_{p_2b}^\nu\right),\nonumber\\
&&\frac{1}{\sqrt{2}}\left(\hat{l}_{p_1a}^\mu \hat{k}_{p_2b}^{\nu} + \hat{k}_{p_1a}^{\mu}\hat{l}_{p_2b}^\nu\right),\nonumber\\
&&\left.\frac{1}{\sqrt{2}}\left(\hat{l}_{p_1a}^{*\mu} \hat{k}_{p_2b}^{\nu} + \hat{k}_{p_1a}^{\mu}\hat{l}_{p_2b}^{*\nu}\right)\right]\nonumber\\
&\otimes&\left.
\left[\hat{l}_q^\alpha,\ \hat{l}_q^{*\alpha},\ \hat{k}_q^\alpha
\right]
\right\},
\label{symglubasis}
\end{eqnarray} 
which reduces the number of tensor basis elements from 27 to 18. Next, to find the explicit constraints imposed by the $\hat{C}$ and $\hat{P}$ transformations, notice that the polarization vectors of kinds $\hat{l}^\mu,\hat{l}^{*\mu}$ and $\hat{k}^\mu$ satisfy
\begin{eqnarray}
\hat{C}\ \hat{l}^\mu\hat{C}^{-1}&=&\hat{P}\ \hat{l}^\mu\hat{P}^{-1}=-\hat{l}^\mu\nonumber\\
\hat{C}\ \hat{l}^{*\mu}\hat{C}^{-1}&=&-\hat{P}\ \hat{l}^{*\mu}\hat{P}^{-1}=-\hat{l}^{*\mu}\nonumber\\
\hat{C}\ \hat{k}^\mu\hat{C}^{-1}&=&-\hat{P}\ \hat{k}^\mu\hat{P}^{-1}=\hat{k}^\mu.
\label{ExplicitProperties}
\end{eqnarray}
Consider the general case whereby we describe a vertex with $n$ gauge bosons out of which $n_l$ are in the polarization state $\hat{l}^\mu$, $n_{l^*}$ in the polarization state $\hat{l}^{*\mu}$ and $n_k$ in the polarization state $\hat{k}^\mu$. The occupation numbers satisfy $n=n_l+n_{l^*}+n_k$. Then, the corresponding quantum numbers under $\hat{C}$ and $\hat{P}$ of the tensor structures in Eq.~(\ref{symglubasis}) are given by the eigenvalues~\cite{Papanyan:1971cv}
\begin{eqnarray}
C_{\mbox{\tiny{tensor}}} &=& (-1)^{n_{l}+n_{l^*}}\nonumber\\
P_{\mbox{\tiny{tensor}}} &=& (-1)^{n_{l}+n_k}
.
\label{eigentensor}
\end{eqnarray}
To preserve invariance under $CP$, the above requires that
the corresponding coefficients transform under $\hat{C}$
and $\hat{P}$ with eigenvalues
\begin{eqnarray}
C_{\mbox{\tiny{coeff}}} &=& (-1)^{n_k}\nonumber\\
P_{\mbox{\tiny{coeff}}} &=& (-1)^{n_{l^*}}
,
\label{eigencoefficients}
\end{eqnarray}
such that the product of eigenvalues for tensor structures and their corresponding coefficients satisfy
\begin{eqnarray}
C_{\mbox{\tiny{tensor}}}P_{\mbox{\tiny{tensor}}}C_{\mbox{\tiny{coeff}}}P_{\mbox{\tiny{coeff}}}&=&(-1)^{n_l+n_{l^*}+n_k}(-1)^{n_l+n_k+n_{l^*}}\nonumber\\
&=&(-1)^{2n}=1.
\label{prodtenscoeff}
\end{eqnarray}
Referring to the coefficients of the $i$-th tensor structure ($i=1\ldots 18$), including their transformation properties under $CP$ ($C,P=\pm$) as
\begin{eqnarray*}
a_i^{CP},
\end{eqnarray*}
we obtain that for general values of the gauge vector momenta, the two-gluon one-photon vertex can be expressed as
\begin{widetext}
\begin{eqnarray}
\Gamma^{\mu\nu\alpha}_{ab}(p_1,p_2,q)&=&a_1^{++}\hat{l}_{p_1a}^\mu \hat{l}_{p_2b}^\nu \hat{l}_q^\alpha + a_2^{++}\hat{l}_{p_1a}^{*\mu} \hat{l}_{p_2b}^{*\nu}\hat{l}_q^\alpha + a_3^{++}\hat{k}_{p_1a}^\mu \hat{k}_{p_2b}^\nu \hat{l}_q^\alpha\nonumber\\
&+& \frac{a_4^{+-}}{\sqrt{2}}\left(\hat{l}_{p_1a}^\mu \hat{l}_{p_2b}^{*\nu} + \hat{l}_{p_1a}^{*\mu}\hat{l}_{p_2b}^\nu\right)\hat{l}_q^\alpha + \frac{a_5^{-+}}{\sqrt{2}}\left(\hat{l}_{p_1a}^\mu \hat{k}_{p_2b}^{\nu} + \hat{k}_{p_1a}^{\mu}\hat{l}_{p_2b}^\nu\right)\hat{l}_q^\alpha + \frac{a_6^{--}}{\sqrt{2}}\left(\hat{l}_{p_1a}^{*\mu} \hat{k}_{p_2b}^{\nu} + \hat{k}_{p_1a}^{\mu}\hat{l}_{p_2b}^{*\nu}\right)\hat{l}_q^\alpha\nonumber\\
&+&a_7^{+-}\hat{l}_{p_1a}^\mu \hat{l}_{p_2b}^\nu \hat{l}_q^{*\alpha} + a_8^{+-}\hat{l}_{p_1a}^{*\mu} \hat{l}_{p_2b}^{*\nu}\hat{l}_q^{*\alpha} + a_9^{+-}\hat{k}_{p_1a}^\mu \hat{k}_{p_2b}^\nu \hat{l}_q^{*\alpha}\nonumber\\
&+& \frac{a_{10}^{++}}{\sqrt{2}}\left(\hat{l}_{p_1a}^\mu \hat{l}_{p_2b}^{*\nu} + \hat{l}_{p_1a}^{*\mu}\hat{l}_{p_2b}^\nu\right)\hat{l}_q^{*\alpha} + \frac{a_{11}^{--}}{\sqrt{2}}\left(\hat{l}_{p_1a}^\mu \hat{k}_{p_2b}^{\nu} + \hat{k}_{p_1a}^{\mu}\hat{l}_{p_2b}^\nu\right)\hat{l}_q^{*\alpha} + \frac{a_{12}^{-+}}{\sqrt{2}}\left(\hat{l}_{p_1a}^{*\mu} \hat{k}_{p_2b}^{\nu} + \hat{k}_{p_1a}^{\mu}\hat{l}_{p_2b}^{*\nu}\right)\hat{l}_q^{*\alpha}\nonumber\\
&+&a_{13}^{-+}\hat{l}_{p_1a}^\mu \hat{l}_{p_2b}^\nu \hat{k}_q^\alpha + a_{14}^{-+}\hat{l}_{p_1a}^{*\mu} \hat{l}_{p_2b}^{*\nu}\hat{k}_q^\alpha + a_{15}^{-+}\hat{k}_{p_1a}^\mu \hat{k}_{p_2b}^\nu \hat{k}_q^\alpha\nonumber\\
&+& \frac{a_{16}^{--}}{\sqrt{2}}\left(\hat{l}_{p_1a}^\mu \hat{l}_{p_2b}^{*\nu} + \hat{l}_{p_1a}^{*\mu}\hat{l}_{p_2b}^\nu\right)\hat{k}_q^{\alpha} + \frac{a_{17}^{++}}{\sqrt{2}}\left(\hat{l}_{p_1a}^\mu \hat{k}_{p_2b}^{\nu} + \hat{k}_{p_1a}^{\mu}\hat{l}_{p_2b}^\nu\right)\hat{k}_q^{\alpha} + \frac{a_{18}^{+-}}{\sqrt{2}}\left(\hat{l}_{p_1a}^{*\mu} \hat{k}_{p_2b}^{\nu} + \hat{k}_{p_1a}^{\mu}\hat{l}_{p_2b}^{*\nu}\right)\hat{k}_q^{\alpha}.
\label{generralvertex}
\end{eqnarray}
\end{widetext}
Notice that in order to write the coefficients $a_i$, $i=1,\ldots, 18$, we have to our disposal Lorentz scalars also with definite properties under $C$ and $P$. Expressing any of the momentum vectors $p_1^\mu, p_2^{\mu}, q^\mu$ as $p_m^\mu$, $m=1,2,3$, the available Lorentz scalars are
\begin{eqnarray}
S_{1\ mn}^{++}&=&(p_m\cdot p_n)\nonumber\\
S_{2\ mn}^{++}&=&(p_m\cdot p_n)_\perp\nonumber\\
S_{3\ mn}^{-+}&=&p_m^\mu\hat{F}_{\mu\nu}p_n^\nu\nonumber\\
S_{4\ mn}^{--}&=&p_m^\mu\hat{F}^*_{\mu\nu}p_n^\nu,
\label{possiblescalars}
\end{eqnarray}
where the notation $S_{j\ mn}^{CP}$, emphasizes the properties under the $C$ and $P$ transformations, respectively, of the corresponding Lorentz scalar of the kind $j=1,\ldots, 4$.

We emphasize that the expression in Eq.~(\ref{generralvertex}) is valid for arbitrary gauge bosons momenta. We now restrict the analysis to the case of on-shell gauge bosons. As previously discussed, in this case we know that
\begin{eqnarray}
k_{p_1}^\mu &\to& p_1^\mu\nonumber\\
k_{p_2}^\mu &\to& p_2^\mu\nonumber\\
k_{q}^\mu &\to& q^\mu.
\label{becomes}
\end{eqnarray}
Therefore, we are required to perform the replacements in Eq.~(\ref{becomes}) into the right-hand side of Eq.~(\ref{generralvertex}). As a consequence,  when computing a matrix element involving this vertex, and thus when contracting with the physical polarization vectors of real gauge bosons, any term where there is a polarization vector of the $k$-kind drops out and thus, the effective vertex becomes
\begin{eqnarray}
\Gamma^{\mu\nu\alpha}_{ab}(p_1,p_2,q)_{\mbox{\tiny{on-shell}}}&=&a_1^{++}\hat{l}_{p_1a}^\mu \hat{l}_{p_2b}^\nu \hat{l}_q^\alpha + a_2^{++}\hat{l}_{p_1a}^{*\mu} \hat{l}_{p_2b}^{*\nu}\hat{l}_q^\alpha\nonumber\\
&+&
\frac{a_4^{+-}}{\sqrt{2}}\left(\hat{l}_{p_1a}^\mu \hat{l}_{p_2b}^{*\nu} + \hat{l}_{p_1a}^{*\mu}\hat{l}_{p_2b}^\nu\right)\hat{l}_q^\alpha \nonumber\\
&+&a_7^{+-}\hat{l}_{p_1a}^\mu \hat{l}_{p_2b}^\nu \hat{l}_q^{*\alpha} + a_8^{+-}\hat{l}_{p_1a}^{*\mu} \hat{l}_{p_2b}^{*\nu}\hat{l}_q^{*\alpha}\nonumber\\
&+&\frac{a_{10}^{++}}{\sqrt{2}}\left(\hat{l}_{p_1a}^\mu \hat{l}_{p_2b}^{*\nu} + \hat{l}_{p_1a}^{*\mu}\hat{l}_{p_2b}^\nu\right)\hat{l}_q^{*\alpha}.
\label{neweffvert}
\end{eqnarray}
A further simplification can be made by noticing that for on-shell gauge bosons, conservation of energy-momentum 
\begin{eqnarray}
\omega_q&=&\omega_{p_1}+\omega_{p_2}\nonumber\\
\Vec{q}&=&\Vec{p}_1+\Vec{p}_2
\label{conservation}
\end{eqnarray}
requires that the gauge bosons are colinear and therefore 
\begin{eqnarray}
p_1^\mu &=& \left(\frac{\omega_{p_1}}{\omega_q}\right)q^\mu\nonumber\\
p_2^\mu &=& \left(\frac{\omega_{p_2}}{\omega_q}\right)q^\mu,
\label{proportionalenergies}
\end{eqnarray}
where we chose to express the gluon momenta in terms of the photon momentum. As a second consequence of dealing with on-shell gauge bosons, we notice that the available scalars to express the coefficients of the tensor structures become
\begin{eqnarray}
S_{1\ mn}^{++}&\to&\left(\frac{\omega_m}{\omega_q}\right)q_\mu\left(\frac{\omega_n}{\omega_q}\right)q^\mu=\left(\frac{\omega_m\omega_n}{\omega_q^2}\right)q^2=0\nonumber\\
S_{2\ mn}^{++}&\to&\left(\frac{\omega_m}{\omega_q}\right)q_\mu^\perp\left(\frac{\omega_n}{\omega_q}\right)q^\mu_\perp=\left(\frac{\omega_m\omega_n}{\omega_q^2}\right)q_\perp^2\nonumber\\
S_{3\ mn}^{-+}&\to&\left(\frac{\omega_m\omega_n}{\omega_q^2}\right)q^\mu\hat{F}_{\mu\nu}q^\nu=0\nonumber\\
S_{4\ mn}^{--}&\to&\left(\frac{\omega_m\omega_n}{\omega_q^2}\right)q^\mu\hat{F}^*_{\mu\nu}q^\nu=0,
\label{possiblescalarsonshell}
\end{eqnarray}
which means that all Lorentz structures, odd with respect to either or both $\hat{C}$ and $\hat{P}$, are not available to express the on-shell effective vertex. Notice that the dependence on $q_\perp^2$ of the surviving scalar does not necessarily mean that the coefficients are to be only proportional to $q_\perp^2$, since this scalar can appear, for example, as an argument of an exponential function in the explicit one-loop calculation. However, notice also that when the the terms that make up the coefficients happen to be proportional to this scalar, they will in turn be proportional to $\sin^2\theta$, where $\theta$ is the angle between the direction of the photon propagation and the direction of the magnetic field. This means that for these kind of terms, a preferred direction of photon propagation is achieved for $\theta = \pi/2$. In fact, as we explicitly show in the Appendix, for the one-loop calculation a large fraction of the terms that contribute to the coefficients show this $\sin\theta$ dependence. 

Putting all this together we can finally write
\begin{eqnarray}
\!\!\!\!\Gamma^{\mu\nu\alpha}_{ab}(p_1,p_2,q)_{\mbox{\tiny{on-shell}}}&=&a_1^{++}\hat{l}_{p_1a}^\mu \hat{l}_{p_2b}^\nu \hat{l}_q^\alpha + a_2^{++}\hat{l}_{p_1a}^{*\mu} \hat{l}_{p_2b}^{*\nu}\hat{l}_q^\alpha\nonumber\\
&+&\frac{a_{10}^{++}}{\sqrt{2}}\left(\hat{l}_{p_1a}^\mu \hat{l}_{p_2b}^{*\nu} + \hat{l}_{p_1a}^{*\mu}\hat{l}_{p_2b}^\nu\right)\hat{l}_q^{*\alpha},
\label{fineffvertex}
\end{eqnarray}
which shows that only three independent tensor structures, together with their corresponding coefficients, are needed to span the on-shell effective two-gluon one-photon vertex. Given that the basis vectors are normalized and are also mutually orthogonal, it is an easy task to compute the coefficients $a_1^{++}$, $a_2^{++}$ and $a_{10}^{++}$ by projecting the vertex $\Gamma^{\mu\nu\alpha}_{ab}$ onto each of the tensors that make up the basis. We provide an explicit example of this procedure in the Appendix, starting from Eq.~(\ref{firstofeqsapp}), for the one-loop calculation of the coefficient $a_1^{++}$.

\section{One-loop approximation for the effective two-gluon one-photon vertex in the intermediate field strength regime}\label{III}

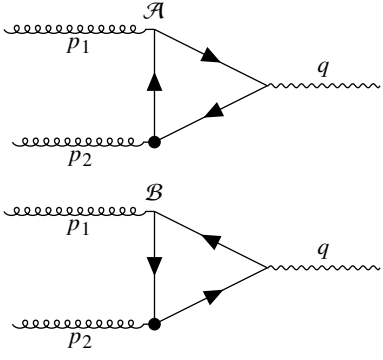
\begin{figure}[t]
\centering
\begin{tikzpicture}
\begin{feynhand}
\vertex (a) at (0,0.75); \vertex  (b) at (0,-0.75) {};  \vertex [dot] (c) at (2,0.75); \vertex [dot] (d) at (2,-0.75) {}; \vertex (e) at (3.5,0); \vertex (f) at (5,0); 
\vertex  (g) at (2,1) {$\mathcal{A}$};
\propag [glu] (a) to [edge label' =$p_1$] (c);
\propag [glu] (b) to [edge label' =$p_2$] (d) ;
\propag [fer] (d) to  (c);
\propag [fer] (c) to  (e);
\propag [fer] (e) to  (d);
\propag [bos] (e) to [edge label =$q$] (f);
\end{feynhand}
\end{tikzpicture}
\begin{tikzpicture}
\begin{feynhand}
\vertex (a) at (0,0.75); \vertex  (b) at (0,-0.75) {};  \vertex  (c) at (2,0.75) ;
\vertex  (g) at (2,1) {$\mathcal{B}$}; \vertex [dot] (d) at (2,-0.75) {}; \vertex (e) at (3.5,0); \vertex (f) at (5,0); 
\propag [glu] (a) to [edge label' =$p_1$] (c);
\propag [glu] (b) to [edge label' =$p_2$] (d) ;
\propag [antfer] (d) to  (c);
\propag [antfer] (c) to  (e);
\propag [antfer] (e) to  (d);
\propag [bos] (e) to [edge label =$q$] (f);
\end{feynhand}
\end{tikzpicture}
\caption{One-loop diagrams contributing to the two-gluon one-photon vertex. Diagram $\mathcal{B}$ represents the charge conjugate of diagram $\mathcal{A}$. The four-momentum vectors are chosen such that $q=p_1+p_2$.}
\label{fig:Fusion}
\end{figure}

At leading order in the strong $\alpha_s$ and electromagnetic $\alpha_{em}$ couplings, the vertex to be used to describe a scattering process involving two gluons and a photon, either gluon fusion or splitting, is depicted in Figure~\ref{fig:Fusion}. The amplitude corresponds to the sum of the Feynman diagrams represented as two fermion triangles with two gluons and one photon attached to the vertices and the charge in one and the other diagrams flowing in opposite directions. 
Since the magnetic field breaks Lorentz symmetry, the vertex needs to be computed starting from configuration space. Each internal line corresponds to a fermion propagator in the presence of a magnetic field, which can be written as
\begin{eqnarray}
S(x,x')=\Phi(x,x')\int\frac{d^4p}{(2\pi)^4}e^{-ip\cdot(x-x')}S(p),
\label{fermionprop}
\end{eqnarray}
where $\Phi(x,x')$ is the Schwinger's phase factor given by
\begin{eqnarray}
\Phi(x,x')=\exp\left\{iq_f\int_{x'}^x\! d\xi^\mu\left[A^\mu+\frac{1}{2}F_{\mu\nu}(\xi-x')^\nu\right]\right\}\!,
\label{schwingerfactor}
\end{eqnarray}
where $q_f$ is the charge of the quark with flavor $f$. The translationally invariant part of the propagator can be written, using Schwinger's proper time representation, as
\begin{eqnarray}
    S(p)&=&\int_0^\infty \frac{ds}{\cos(q_fBs)}e^{is\left(p_\parallel^2+p_\perp^2\frac{\tan(q_fBs)}{q_fBs}-m_f^2+i\epsilon\right)}\nonumber\\
&\times&\left[
e^{iq_fBs\Sigma_3}\left(m_f +\slashed{p}_\parallel\right) + \frac{\slashed{p}_\perp}{\cos(q_fBs)}
\right],\label{fermionpropagatormomentumspace}
\end{eqnarray}
where $m_f$ is the mass of the quark with flavor $f$ and $\Sigma_3=i\gamma_1\gamma_2$. The explicit expression for the sum of the diagrams in Fig.~\ref{fig:Fusion} is written as
\begin{eqnarray}
   \Gamma^{\mu\nu\alpha}_{ab}&=&-ig^2 q_f\int \!d^4xd^4yd^4z\int\!\frac{d^4 r_1}{(2\pi)^4}
   \frac{d^4 r_2}{(2\pi)^4}\frac{d^4 r_3}{(2\pi)^4}\nonumber\\
   &\times&e^{-i r_3\cdot (y-x)}e^{-i r_2\cdot (x-z)}e^{-i r_1\cdot (z-y)}e^{-i p_1\cdot z}e^{-i p_2\cdot y}e^{i q\cdot x}\nonumber\\
   &\times&
   \Big\{
   {\mbox{Tr}}\left[\gamma_\alpha S(r_2) \gamma_\mu t_a S(r_1)\gamma_\nu t_b S(r_3) \right]\Phi(x,y,z,x)\nonumber\\
   &+&
   {\mbox{Tr}}\left[\gamma_\alpha  S(r_3)\gamma_\nu t_b S(r_1) \gamma_\mu t_a S(r_2) \right]\Phi(x,z,y,x)
   \Big\},\nonumber\\
   \label{amplitude1}  
\end{eqnarray}
where
\begin{eqnarray*}
    \Phi(x,y,z,x) & = & \Phi(x,y)\Phi(y,z)\Phi(z,x)\nonumber\\
    \Phi(x,z,y,x) & = & \Phi(x,z)\Phi(z,y)\Phi(y,x)=\Phi^*(x,y,z,x)
\end{eqnarray*}
and $g$ is the quark-gluon coupling, $t^a=\lambda^a/2,\ t^b=\lambda^b/2$ with $\lambda^a$ and $\lambda^b$ being Gell-Mann matrices. To describe a constant magnetic field that points in the $z$-direction, the vector potential $A^\mu$ can be chosen in the symmetric gauge,
\begin{eqnarray}
A^\mu=\frac{B}{2}(0,-y,x,0), 
\end{eqnarray}
so that, from Eq.~(\ref{schwingerfactor}), the product of Schwinger phases is given by
\begin{eqnarray}
\!\!\!\Phi(x,y)\Phi(y,z)\Phi(z,x)&=&e^{-i\frac{|q_fB|}{2}(z-x)_m\hat{F}_{mj}(x-y)_j},\nonumber\\ 
\!\!\!\Phi(x,z)\Phi(z,y)\Phi(y,x) & = & e^{-i\frac{|q_fB|}{2}(y-x)_m\hat{F}_{mj}(x-z)_j}
\end{eqnarray}
where the indices $m,j=1,2$. Notice that the product of Schwinger phases is gauge invariant~\cite{Ayala:2020muk}. The integrals over the space-time variables $x,y,z$ can be performed straightforward after identifying Dirac delta functions, yielding
\begin{eqnarray}
\Gamma^{\mu\nu\alpha}_{ab}&=&\!\!-i\frac{g^2q_f}{(2\pi^2)(q_fB)^2}{\mbox{Tr}}[t_at_b]
\int\!d^4r_1d^4r_2d^4r_3\nonumber\\
   &\times&\!\!\delta^2(p_1-r_2+r_1)_\parallel\delta^2(p_2-r_1+r_3)_\parallel\delta^2(q+r_3-r_2)_\parallel\nonumber\\
   &\times&\!\!\delta^2(p_1+p_2-q)_\perp \nonumber\\
   &\times&\!\!\Big\{
   {\mbox{Tr}}\left[\gamma_\alpha S(r_2) \gamma_\mu S(r_1) \gamma_\nu S(r_3) \right]\textcolor{black}{\widetilde{\Phi}(p_1,p_2;r_1,r_2,r_3)}\nonumber\\
   &+&\!\!
   {\mbox{Tr}}\left[\gamma_\alpha S(r_3)\gamma_\nu S(r_1) \gamma_\mu S(r_2) \right] \textcolor{black}{\widetilde{\Phi}^*(p_1,p_2;r_1,r_2,r_3)}
   \Big\},\nonumber\\
\end{eqnarray}
where
\begin{eqnarray*}
    \textcolor{black}{\widetilde{\Phi}(p_1,p_2;r_1,r_2,r_3)=e^{\frac{2i}{|q_fB|}(p_1-r_2+r_1)_m\hat{F}_{mj}(p_2+r_3-r_1)_j}}
\end{eqnarray*}
Using the delta functions to perform the integrations over the corresponding momentum components, we get
\begin{eqnarray}
\Gamma^{\mu\nu\alpha}_{ab}&\!\!=\!\!&-i\frac{g^2q_f}{(2\pi^2)(q_fB)^2}{\mbox{Tr}}[t_at_b]
\delta^4(p_1+p_2-q)\nonumber\\
   &\!\!\times\!\!&\int\!d^4r_1d^2r_2d^2r_3e^{\frac{2i}{|q_fB|}(p_1-r_2+r_1)_m\hat{F}_{mj}(p_2+r_3-r_1)_j}\nonumber\\
   &\!\!\times\!\!&\Big\{
   {\mbox{Tr}}\left[\gamma_\alpha S(r_2) \gamma_\mu S(r_1) \gamma_\nu S(r_3) \right]\textcolor{black}{\widetilde{\Phi}(p_1,p_2;r_1,r_2,r_3)}\nonumber\\
   &\!\!+\!\!&
   {\mbox{Tr}}\left[\gamma_\alpha S(r_3)\gamma_\nu S(r_1) \gamma_\mu S(r_2) \right]\textcolor{black}{\widetilde{\Phi}^*(p_1,p_2;r_1,r_2,r_3)}
   \Big\}.\nonumber\\
\end{eqnarray}

To carry out the integrations over the remaining internal momentum components, we now follow the procedure described in Ref.~\cite{Jaber-Urquiza:2023swn}. We first introduce the short-hand notation
\begin{eqnarray}
c_j&\equiv&\cos(q_fBs_j),\nonumber\\
t_j&\equiv&\tan(q_fBs_j),\nonumber\\
e_j&\equiv&c_je^{i{\mbox{\small{sign}}(q_fB) }q_fBs_j\Sigma_3},
\end{eqnarray}
with $s_j$ being the Schwinger proper-time parameters associated to the internal fermion lines. The momentum integrations can be computed by standard Gaussian integration. After a lengthy but straightforward calculation, \textcolor{black}{where we take the on-shell limit,} we get
\begin{eqnarray}
    \Gamma^{\mu\nu\alpha}_{ab}&=&-i\frac{g^2q_f^2B\ }{(2\pi^2)}{\mbox{Tr}}[t_at_b]
    \delta^4(p_1+p_2-q)\nonumber\\
    &\times&\int_0^\infty\frac{ds_1ds_2ds_3}{c_1^2c_2^2c_3^2}\left(\frac{1}{t_1t_2t_3-t_1-t_2-t_3}\right)\left(\frac{e^{-ism_f^2}}{s}\right)\nonumber\\
    &\times&
    e^{-\frac{i}{s}\left(s_1s_3\omega_{p_1}^2+s_2s_3\omega_{p_2}^2+s_1s_2\omega_q^2\right)\frac{q_\perp^2}{\omega_q^2}}\nonumber\\
    &\times&e^{-\frac{i}{\omega_q^2}\frac{q_\perp^2}{|q_fB|}\left(\frac{1}{t_1t_2t_3-t_1-t_2-t_3}\right)\left(t_1t_3\omega_{p_1}^2+t_2t_3\omega_{p_2}^2+t_1t_2\omega_q^2\right)}\nonumber\\
    &\times&\sum_{j=1}^{19}\left(T_{{\mathcal{A}}j}^{\mu\nu\alpha}+T_{{\mathcal{B}}j}^{\mu\nu\alpha}\right),
    \label{matelem}
    \end{eqnarray}
where $s=s_1+s_2+s_3$ and the traces over Dirac space corresponding to diagrams ${\mathcal{A}}$ and ${\mathcal{B}}$ of Fig.~\ref{fig:Fusion}, and that depend on the Schwinger parameters as well as on the photon and gluons momenta are given by
\begin{widetext}
\begin{eqnarray}
T_{{\mathcal{A}}1}^{\mu\nu\alpha}+T_{{\mathcal{B}}1}^{\mu\nu\alpha}&=&{\mbox{Tr}}[\gamma^\mu \slashed{\mathcal{A}}_a\gamma^\alpha\slashed{\mathcal{A}}_b\gamma^\nu\slashed{\mathcal{A}}_c]
+
{\mbox{Tr}}[\gamma^\mu \slashed{\mathcal{B}}_c\gamma^\nu\slashed{\mathcal{B}}_b\gamma^\alpha\slashed{\mathcal{B}}_a],
\nonumber\\
T_{{\mathcal{A}}2}^{\mu\nu\alpha}+T_{{\mathcal{B}}2}^{\mu\nu\alpha}&=&m_f^2\left\{{\mbox{Tr}}[\gamma^\mu e_1\gamma^\alpha e_2\gamma^\nu\slashed{\mathcal{A}}_c]
+
{\mbox{Tr}}[\gamma^\mu \slashed{\mathcal{B}}_c\gamma^\nu e_2\gamma^\alpha e_1]\right\},\nonumber\\
T_{{\mathcal{A}}3}^{\mu\nu\alpha}+T_{{\mathcal{B}}3}^{\mu\nu\alpha}&=&m_f^2\left\{{\mbox{Tr}}[\gamma^\mu e_1\gamma^\alpha \slashed{\mathcal{A}}_b\gamma^\nu e_3]
+
{\mbox{Tr}}[\gamma^\mu e_3\gamma^\nu \slashed{\mathcal{B}}_b\gamma^\alpha e_1]\right\},\nonumber\\
T_{{\mathcal{A}}4}^{\mu\nu\alpha}+T_{{\mathcal{B}}4}^{\mu\nu\alpha}&=&m_f^2\left\{{\mbox{Tr}}[\gamma^\mu \slashed{\mathcal{A}}_a\gamma^\alpha e_2\gamma^\nu e_3]
+
{\mbox{Tr}}[\gamma^\mu e_3\gamma^\nu e_2\gamma^\alpha \slashed{\mathcal{B}}_a]\right\},\nonumber\\
T_{{\mathcal{A}}5}^{\mu\nu\alpha}+T_{{\mathcal{B}}5}^{\mu\nu\alpha}&=&\frac{i}{s}\left\{{\mbox{Tr}}[\gamma^\mu \slashed{\mathcal{A}}_a\gamma^\alpha e_2\gamma^\nu_\parallel e_3]
+
{\mbox{Tr}}[\gamma^\mu e_3\gamma^\nu_\parallel e_2\gamma^\alpha \slashed{\mathcal{B}}_a]\right\},\nonumber\\
T_{{\mathcal{A}}6}^{\mu\nu\alpha}+T_{{\mathcal{B}}6}^{\mu\nu\alpha}&=&\frac{i}{s}\left\{{\mbox{Tr}}[\gamma^\mu_\parallel e_1\gamma^\alpha \slashed{\mathcal{A}}_b \gamma^\nu e_3]
+
{\mbox{Tr}}[\gamma^\mu_\parallel e_3\gamma^\nu \slashed{\mathcal{B}}_b\gamma^\alpha  e_1]\right\},\nonumber\\
T_{{\mathcal{A}}7}^{\mu\nu\alpha}+T_{{\mathcal{B}}7}^{\mu\nu\alpha}&=&\frac{i}{s}\left\{{\mbox{Tr}}[\gamma^\mu e_1\gamma^\alpha_\parallel e_2 \gamma^\nu \slashed{\mathcal{A}}_c]
+
{\mbox{Tr}}[\gamma^\mu \slashed{\mathcal{B}}_c\gamma^\nu e_2 \gamma^\alpha_\parallel  e_1]\right\},\nonumber\\
T_{{\mathcal{A}}8}^{\mu\nu\alpha}+T_{{\mathcal{B}}8}^{\mu\nu\alpha}&=&-\frac{i}{s}\left\{{\mbox{Tr}}[\gamma^\mu \slashed{\mathcal{A}}_a \gamma^\alpha e_2 \gamma^\nu e_3 ]
+
{\mbox{Tr}}[\gamma^\mu e_3\gamma^\nu e_2 \gamma^\alpha \slashed{\mathcal{B}}_a]\right\},\nonumber\\
T_{{\mathcal{A}}9}^{\mu\nu\alpha}+T_{{\mathcal{B}}9}^{\mu\nu\alpha}&=&-\frac{i}{s}\left\{{\mbox{Tr}}[\gamma^\mu e_1 \gamma^\alpha \slashed{\mathcal{A}}_b\gamma^\nu e_3 ]
+
{\mbox{Tr}}[\gamma^\mu e_3\gamma^\nu \slashed{\mathcal{B}}_b\gamma^\alpha  e_1]\right\},\nonumber\\
T_{{\mathcal{A}}10}^{\mu\nu\alpha}+T_{{\mathcal{B}}10}^{\mu\nu\alpha}&=&-\frac{i}{s}\left\{{\mbox{Tr}}[\gamma^\mu e_1 \gamma^\alpha e_2\gamma^\nu\slashed{\mathcal{A}}_c]
+
{\mbox{Tr}}[\gamma^\mu \slashed{\mathcal{B}}_c \gamma^\nu e_2\gamma^\alpha  e_1]\right\},\nonumber\\
T_{{\mathcal{A}}11}^{\mu\nu\alpha}+T_{{\mathcal{B}}11}^{\mu\nu\alpha}&=&\frac{iq_fB}{(t_1t_2t_3-t_1-t_2-t_3)}\left\{{\mbox{Tr}}[\gamma^\mu  \slashed{\mathcal{A}}_a\gamma^\alpha \gamma^\nu]
+
{\mbox{Tr}}[\gamma^\mu \gamma^\nu\gamma^\alpha  \slashed{\mathcal{B}}_a]\right\},\nonumber\\
T_{{\mathcal{A}}12}^{\mu\nu\alpha}+T_{{\mathcal{B}}12}^{\mu\nu\alpha}&=&\frac{iq_fB}{(t_1t_2t_3-t_1-t_2-t_3)}\left\{{\mbox{Tr}}[\gamma^\mu  \gamma^\alpha\slashed{\mathcal{A}}_b\gamma^\nu]
+
{\mbox{Tr}}[\gamma^\mu \gamma^\nu\slashed{\mathcal{B}}_b\gamma^\alpha]\right\},\nonumber\\
T_{{\mathcal{A}}13}^{\mu\nu\alpha}+T_{{\mathcal{B}}13}^{\mu\nu\alpha}&=&\frac{iq_fB}{(t_1t_2t_3-t_1-t_2-t_3)}\left\{{\mbox{Tr}}[\gamma^\mu  \gamma^\alpha\gamma^\nu\slashed{\mathcal{A}}_c]
+
{\mbox{Tr}}[\gamma^\mu\slashed{\mathcal{B}}_c \gamma^\nu\gamma^\alpha]\right\},\nonumber\\
T_{{\mathcal{A}}14}^{\mu\nu\alpha}+T_{{\mathcal{B}}14}^{\mu\nu\alpha}&=&-\frac{iq_fB}{(t_1t_2t_3-t_1-t_2-t_3)}\left\{{\mbox{Tr}}[\gamma^\mu\slashed{\mathcal{A}}_a\gamma^\alpha\gamma^\nu_\perp]
+
{\mbox{Tr}}[\gamma^\mu\gamma^\nu_\perp\gamma^\alpha\slashed{\mathcal{B}}_a]\right\},\nonumber\\
T_{{\mathcal{A}}15}^{\mu\nu\alpha}+T_{{\mathcal{B}}15}^{\mu\nu\alpha}&=&-\frac{iq_fB}{(t_1t_2t_3-t_1-t_2-t_3)}\left\{{\mbox{Tr}}[\gamma^\mu_\perp\gamma^\alpha\slashed{\mathcal{A}}_b\gamma^\nu]
+
{\mbox{Tr}}[\gamma^\mu_\perp\gamma^\nu\slashed{\mathcal{B}}_b\gamma^\alpha]\right\},\nonumber\\
T_{{\mathcal{A}}16}^{\mu\nu\alpha}+T_{{\mathcal{B}}16}^{\mu\nu\alpha}&=&-\frac{iq_fB}{(t_1t_2t_3-t_1-t_2-t_3)}\left\{{\mbox{Tr}}[\gamma^\mu\gamma^\alpha_\perp\gamma^\nu\slashed{\mathcal{A}}_c]
+
{\mbox{Tr}}[\gamma^\mu\slashed{\mathcal{B}}_c\gamma^\nu\gamma^\alpha_\perp]\right\},\nonumber\\
T_{{\mathcal{A}}17}^{\mu\nu\alpha}+T_{{\mathcal{B}}17}^{\mu\nu\alpha}&=&\frac{iq_fB\ t_1}{2(t_1t_2t_3-t_1-t_2-t_3)}\left\{{\mbox{Tr}}[\gamma^\mu\slashed{\mathcal{A}}_a\gamma^\alpha\gamma^\beta_\perp\gamma^\nu\gamma^\sigma_\perp]\hat{F}_{\beta\sigma}
+
{\mbox{Tr}}[\gamma^\mu\gamma^\sigma_\perp\gamma^\nu\gamma^\beta_\perp\gamma^\alpha\slashed{\mathcal{B}}_a]\hat{F}_{\sigma\beta}\right\},
\nonumber
\end{eqnarray}
\begin{eqnarray}
T_{{\mathcal{A}}18}^{\mu\nu\alpha}+T_{{\mathcal{B}}18}^{\mu\nu\alpha}&=&-\frac{iq_fB\ t_2}{2(t_1t_2t_3-t_1-t_2-t_3)}\left\{{\mbox{Tr}}[\gamma^\mu\gamma^\beta_\perp\gamma^\alpha\slashed{\mathcal{A}}_b\gamma^\nu\gamma^\sigma_\perp]\hat{F}_{\beta\sigma}
+
{\mbox{Tr}}[\gamma^\mu\gamma^\sigma_\perp\gamma^\nu\slashed{\mathcal{B}}_b\gamma^\alpha\gamma^\beta_\perp]\hat{F}_{\sigma\beta}\right\},\nonumber\\
T_{{\mathcal{A}}19}^{\mu\nu\alpha}+T_{{\mathcal{B}}19}^{\mu\nu\alpha}&=&\frac{iq_fB\ t_3}{2(t_1t_2t_3-t_1-t_2-t_3)}\left\{{\mbox{Tr}}[\gamma^\mu\gamma^\beta_\perp\gamma^\alpha\gamma^\sigma_\perp\gamma^\nu\slashed{\mathcal{A}}_c]\hat{F}_{\beta\sigma}
+
{\mbox{Tr}}[\gamma^\mu\slashed{\mathcal{B}}_c\gamma^\nu\gamma^\sigma_\perp\gamma^\alpha\gamma^\beta_\perp]\hat{F}_{\sigma\beta}\right\},\nonumber\\
\label{expressions1}
\slashed{\mathcal{A}}_a&=&-\left(\frac{s_3\omega_{p_1}+s_2\omega_q}{s\omega_q}\right)\slashed{q}_\parallel e_1 + \frac{(t_3\omega_{p_1}+t_2\omega_q)\slashed{q}_\perp - t_2t_3\omega_{p_2}\gamma^\sigma\hat{F}_{\sigma\beta}q_\perp^\beta}{(t_1t_2t_3-t_1-t_2-t_3)\omega_q}
\nonumber\\
\slashed{\mathcal{A}}_b&=&\left(\frac{s_1\omega_{q}+s_3\omega_{p_2}}{s\omega_q}\right)\slashed{q}_\parallel e_2 - \frac{(t_3\omega_{p_2}+t_1\omega_q)\slashed{q}_\perp + t_1t_3\omega_{p_1}\gamma^\sigma\hat{F}_{\sigma\beta}q_\perp^\beta}{(t_1t_2t_3-t_1-t_2-t_3)\omega_q}
\nonumber\\
\slashed{\mathcal{A}}_c&=&\left(\frac{s_1\omega_{p_1}-s_2\omega_{p_2}}{s\omega_q}\right)\slashed{q}_\parallel e_3 + \frac{(-t_1\omega_{p_1}+t_2\omega_{p_2})\slashed{q}_\perp + t_1t_3\omega_{q}\gamma^\sigma\hat{F}_{\sigma\beta}q_\perp^\beta}{(t_1t_2t_3-t_1-t_2-t_3)\omega_q}
\nonumber\\
\slashed{\mathcal{B}}_a&=&\left(\frac{s_3\omega_{p_1}+s_2\omega_{q}}{s\omega_q}\right)\slashed{q}_\parallel e_1 - \frac{(t_3\omega_{p_1}+t_2\omega_{q})\slashed{q}_\perp + t_2t_3\omega_{p_2}\gamma^\sigma\hat{F}_{\sigma\beta}q_\perp^\beta}{(t_1t_2t_3-t_1-t_2-t_3)\omega_q}\nonumber\\
\slashed{\mathcal{B}}_b&=&-\left(\frac{s_1\omega_{q}+s_3\omega_{p_2}}{s\omega_q}\right)\slashed{q}_\parallel e_2 + \frac{(t_3\omega_{p_2}+t_1\omega_{q})\slashed{q}_\perp - t_1t_3\omega_{p_1}\gamma^\sigma\hat{F}_{\sigma\beta}q_\perp^\beta}{(t_1t_2t_3-t_1-t_2-t_3)\omega_q}\nonumber\\
\slashed{\mathcal{B}}_c&=&-\left(\frac{s_1\omega_{p_1}-s_2\omega_{p_2}}{s\omega_q}\right)\slashed{q}_\parallel e_3 + \frac{(t_1\omega_{p_1}-t_2\omega_{p_2})\slashed{q}_\perp + t_1t_3\omega_{q}\gamma^\sigma\hat{F}_{\sigma\beta}q_\perp^\beta}{(t_1t_2t_3-t_1-t_2-t_3)\omega_q}.
\end{eqnarray}
\end{widetext}
The above equations represent the exact one-loop result for the two-gluon one-photon vertex in the presence of a constant magnetic field of arbitrary strength. Its large field approximation has been explored in Refs.~\cite{Ayala:2017vex,Ayala:2019jey,Ayala:2022zhu}. However, when recalling that the field strength reaches its peak intensity at the very early stages of a heavy-ion collision and moreover that this is at most a few times the pion mass squared, the large field
\begin{figure}[b]
    \centering
    \begin{tikzpicture}
        \draw [-{Triangle}, thick] (-1,0) -- (5.5,0);
        \draw [-{Triangle}, thick] (0,-1) -- (0,3);
        \draw [blue, ultra thick] (0,0) -- (0.7,0);
        \draw [blue, ultra thick] (1.3,0) -- (1.7,0);
        \draw [blue, ultra thick] (2.3,0) -- (2.7,0);
        \draw [blue, ultra thick,->] (2.3,0) -- (2.6,0);
        \draw [blue, ultra thick] (3.3,0) -- (3.7,0);
        \draw [blue, ultra thick] (4.3,0) -- (4.5,0);
        \draw [blue, ultra thick] (0.7,0) arc (0:180:-0.3);
        \draw [blue, ultra thick] (1.7,0) arc (0:180:-0.3);
        \draw [blue, ultra thick] (2.7,0) arc (0:180:-0.3);
        \draw [blue, ultra thick] (3.7,0) arc (0:180:-0.3);
        \draw [blue, ultra thick] (0,0) -- (3.897,2.25);
        \draw [blue, ultra thick,->]  (3.897,2.25)--(1.9485,1.125);
        \draw [blue, ultra thick,->] (4.5,0) arc (0:15:4.5);
        \draw [blue, ultra thick] (4.5,0) arc (0:30:4.5);
        \filldraw [red] (1,0) circle (2.5pt);
        \filldraw [red] (2,0) circle (2.5pt);
        \filldraw [red] (3,0) circle (2.5pt);
        \filldraw [red] (4,0) circle (2.5pt);
        \node [below] at (4.5,0) {$R$};
        \node [anchor=west] at (4.5,1.25) {$C_1$};
        \node [anchor=east] at (2.5,1.7) {$C_2$};
        \node [anchor=east] at (5.5,2.8) {$x$};
        \draw [black, thick] (5.15,2.65) -- (5.4,2.65);  
         \draw [black, thick] (5.15,2.65) -- (5.15,2.9);  
        \node [above] at (1.4,0.03) {$\theta=\pi/6$};
        \draw [thick, ->] (0.75,0) arc (0:30:0.75);
    \end{tikzpicture}
    \caption{Integration contour in the complex $x$-plane used in Eqs.~\eqref{eq:int_form}. The red dots along the horizontal axis represents the poles of the function $\csc^n (x)$ and they are circumvented using Cauchy's principal value. The path is counterclockwise and this is indicated by the arrows along the integration contour.}
 \label{fig:Int_cont}
\end{figure}
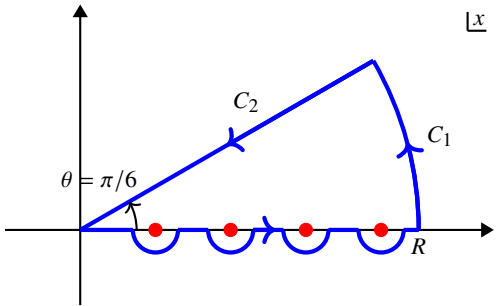
approximation breaks down for photon transverse momenta $q_\perp\gtrsim 200 - 300$ MeV. In this case and in order to compare the results with the measured photon yield and $\nu_2$, which are reported for $q_\perp\gtrsim 500$ MeV, it makes more sense to explore the limit when $q_\perp^2/|q_fB|> 1$. Since on the other hand, the active quark mass species are the light ones, the regime that is worth exploring is such that the field strength and the quark masses satisfy $m_f^2<|q_fB|$. This is the intermediate field strength regime, where the hierarchy of scales becomes $m_f^2<|q_fB|<q_\perp^2$. To carry out the calculation, we perform the change of variable $s_i \equiv s v_i$, $i=1,2,3$, with $v_1 + v_2 + v_3 = 1$, which means that we can express one of these variables in terms of the other two, for instance $v_3=1-v_1-v_2$. The remaining integrals to perform are over the variables $s,v_1,v_2$ with domains $0\leq s \leq \infty$, $0\leq v_1 \leq 1$ and $0\leq v_2\leq 1-v_1$. 

In general, the terms in the expression for $\Gamma_{ab}^{\mu\nu\alpha}$ have the form
\begin{equation}
    \sum_{j=1}^{19} \int ds \, dv_1 dv_2\: K \,e^{i\text{Arg}} T^{\mu\nu\alpha}_j,
\end{equation}
where
\begin{IEEEeqnarray}{rCl}
    K & = & -i\frac{g^2q_f^2B\ }{(2\pi^2)}{\Tr}[t_at_b] \nonumber\\
    &\times&\frac{s}{c_1^2c_2^2c_3^2}\left(\frac{1}{t_1t_2t_3-t_1-t_2-t_3}\right), \IEEEeqnarraynumspace \\
   \text{Arg} & = &  -s(m_f^2) -s\left(v_1 (1-v_1-v_2)\omega_{p_1}^2 \right.\nonumber\\
   &+& \left. v_2 (1-v_1-v_2)\omega_{p_2}^2+v_1 v_2\omega_q^2\right)\frac{q_\perp^2}{\omega_q^2}\nonumber \\
   &-& \frac{1}{\omega_q^2}\frac{q_\perp^2}{|q_fB|}\left(\frac{1}{t_1t_2t_3-t_1-t_2-t_3}\right)\nonumber \\
   &\times&\left(t_1t_3\omega_{p_1}^2+t_2t_3\omega_{p_2}^2+t_1t_2\omega_q^2\right), \\
   T^{\mu\nu\alpha}_j & = & T_{\mathcal{A}j} ^{\mu\nu\alpha} + T_{\mathcal{B}j} ^{\mu\nu\alpha}.
\end{IEEEeqnarray}
Notice that $K$ can be expressed as
\begin{IEEEeqnarray}{rCl}
    K & = & i\frac{g^2q_f^2B\ }{(2\pi^2)}{\Tr}[t_at_b] s \csc (q_f B s)\sec (q_f B s v_1) \nonumber \\
    &\times& \sec (q_f B s v_2) \sec (q_f B s (1-v_1-v_2)).\IEEEeqnarraynumspace
\end{IEEEeqnarray}
\textcolor{black}{
The region of the integrand that provides the largest contribution in the current regime corresponds to the case in which the dimensionless product $q_fBs$ is small~\cite{Piccinelli:2017yvl,Jaber-Urquiza:2018oex}. 
In this case,} we can expand Arg for small $q_f B\textcolor{black}{s}$. Considering the expansion up to third order, we get
\begin{IEEEeqnarray}{rCl}
    \text{Arg} & = & - m_f^2 s+\frac{q_\perp^2 |q_f B|^2}{3 \omega_q^2} s^3 \left(v_1^4 \omega_{p_1}^2+2 v_1^3 \left((v_2-1) \omega_{p_1}^2\right. \right. \nonumber \\
    & + & \left. v_2 (\omega_2^2-\omega_q^2)\right) + v_1^2 \left((3 v_2^2-3 v_2+1) \omega_{p_1}^2\right.\nonumber \\
    & + & \left. 3 (v_2-1) v_2 (\omega_2^2-\omega_q^2)\right) \nonumber \\
    & + & v_1 v_2 \left(2 v_2^2-3 v_2+1\right) \left(\omega_{p_1}^2+\omega_2^2-\omega_q^2\right)\nonumber\\
    &+&\left.(v_2-1)^2 v_2^2 \omega_2^2\right) \nonumber \\
    & \equiv & - m_f ^2 s +  q_\perp ^2 |q_f B|^2 s^3 F(v_1 , v_2),
\end{IEEEeqnarray}
where
\begin{IEEEeqnarray}{rCl}
    F(v_1,v_2) & = & \frac{1}{3\omega_q^2} \left(v_1^4 \omega_{p_1}^2+2 v_1^3 \left((v_2-1) \omega_{p_1}^2+v_2 (\omega_2^2-\omega_q^2)\right) \right. \nonumber \\
    & + & v_1^2 \left((3 v_2^2-3 v_2+1) \omega_{p_1}^2+3 (v_2-1) v_2 (\omega_2^2-\omega_q^2)\right) \nonumber \\
    & + & v_1 v_2 \left(2 v_2^2-3 v_2+1\right) \left(\omega_{p_1}^2+\omega_2^2-\omega_q^2\right) \nonumber\\
    &+&\left.(v_2-1)^2 v_2^2 \omega_2^2\right).
\end{IEEEeqnarray}
 We can further simplify the above expressions recalling that in the intermediate field regime, terms proportional to $m_f^2$ can be neglected. Defining $s=x/q_f B$, we write
\begin{IEEEeqnarray}{rCl}
    K & = & i\frac{g^2q_f }{16\pi^2}\Tr[t_a t_b]x \csc(x) \nonumber \\
    &\times& \sec(v_1 x) \sec(v_2 x) \sec(x (1-v_1-v_2)), \\
    \text{Arg} & = & \frac{q_\perp^2}{|q_f B|}x^3 F(v_1,v_2).
\end{IEEEeqnarray}
Notice that since the function $K$ is proportional to $\csc(x)$, three kinds of integrals contribute to the  the expression, These can be written as (see the Appendix)
\begin{IEEEeqnarray}{rCl}
\label{eq:int_form}
    J_1 & \equiv & \int dv_1 dv_2 dx\, \csc^2(x) e^{i\text{Arg}} G_1(x,v_1,v_2), \\
    J_2 & \equiv & \int dv_1 dv_2 dx\, \csc^3(x) e^{i\text{Arg}} G_2(x,v_1,v_2), \\
    J_3 & \equiv & \int dv_1 dv_2 dx\, \csc^4(x) e^{i\text{Arg}} G_3(x,v_1,v_2).
\end{IEEEeqnarray}
    The integrands of the functions $J_1$, $J_2$ and $J_3$ contain poles along the real $x$-axis, which  correspond to the poles of $\csc^{n}(x)$. The pole located at $x=0$ is removed because $G_i\to 0$, $i=1,2,3$, as $x\to 0$. The rest of the poles are located at $x = n\pi$, with $n\in \mathbb{N}$. To evaluate these integrals, we resort to the integration contour represented in Fig.~\ref{fig:Int_cont}. We can then evaluate the integrals resorting to the residue theorem
\begin{widetext}
\begin{IEEEeqnarray}{rCl}
    \int_{\mathbb{R}^+ \cup C_1 \cup C_2} dx \csc^{j+1} (x) e^{i\text{Arg}} G_j(x,v_1,v_2) =  \sum_{n=1}^\infty\text{Res}\left(\csc^{j+1} (x) e^{i\text{Arg}} G_j(x,v_1,v_2),n\pi\right),
\end{IEEEeqnarray}
where
\begin{IEEEeqnarray}{rCl}
    \int_{\mathbb{R}^+ \cup C_1 \cup C_2} dx \csc^{j+1} (x) e^{i\text{Arg}} G_j(x,v_1,v_2)
     & = &  \mathcal{P}\int_0 ^\infty dx \csc^{j+1} (x) e^{i\text{Arg}} G_j(x,v_1,v_2)
    + \int_{C_1} dx \csc^{j+1} (x) e{i^\text{Arg}} G_j(x,v_1,v_2) \nonumber \\
    & + & \int_{C_2} dx \csc^{j+1} (x) e^{i\text{Arg}} G_j(x,v_1,v_2).
\end{IEEEeqnarray}
For the integral along $C_1$, we can parametrize $x=R e^{i\theta}$, with $R\gg 1$ and $0\leq\theta \leq\pi/6$. Thus, $\text{Arg} =  \tfrac{q_\perp ^2}{q_f B}R^3 e^{i 3 \theta} F(v_1,v_2) $. For sufficiently large $R$ , the real part of $i$Arg is always negative, and hence $e^{i\text{Arg}} \to 0$ as $R\to\infty$. For the integral along $C_2$, we can parametrize $x= \tau e^{i\pi/6}$. Thus, $i\text{Arg} = - \tfrac{q_\perp ^2}{q_f B}\tau^3 e^{i 3 \pi / 6} F(v_1,v_2) =  -\tfrac{q_\perp ^2}{q_f B}\tau^3 F(v_1,v_2)$. Therefore, the integral over $x$ on the real axis can be written as
\begin{IEEEeqnarray}{rCl}
    \int_0 ^\infty dx \csc^{j+1} (x) e^{i\text{Arg}} G_j(x,v_1,v_2)
    & = & \sum \text{Res}\left(\csc^{j+1} (x) e^{i\text{Arg}} G_j(x,v_1,v_2),n\pi\right)\nonumber\\
    & - &\int_0 ^\infty d\tau \csc^{j+1} (\tau e^{i\pi / 6}) 
    e^{-\tfrac{q_\perp ^2}{q_f B}\tau^3 F(v_1,v_2)} G_j(\tau e^{i\pi / 6}, v_1, v_2).
    \label{Int_x}
\end{IEEEeqnarray}
\end{widetext}
The integral on the right-hand side of Eq.~(\ref{Int_x}) can now be integrated, including integration over the variables $v_1$ and $v_2$, numerically. The integration over the variables $v_1$ and $v_2$ for the first term on the right-hand side of Eq.~(\ref{Int_x}) is of the form
\begin{equation}
    \int dv_1 dv_2 \, \mathcal{F}(v_1,v_2) e^{i \chi \psi(v_1,v_2)},
\end{equation}
where the parameter $\chi = \tfrac{q_\perp^2}{q_f B}$ and the phase $\psi$ has a set of critical points $\Upsilon=(v_1^0,v_2^0)$ where $\nabla \psi(v_1^0,v_2^0) = 0$. Therefore, we can make use of the stationary phase approximation, which gives the asymptotic behaviour of this integral for $\chi\gg 1$, as
\begin{IEEEeqnarray}{rCl}
    \int d^2v\, \mathcal{F}(\mathbf{v}) e^{i \chi \psi(\mathbf{v})} & \approx & \sum_{\vec{v}^0\in \Upsilon}e^{i \chi \psi(\vec{v}_0)}\left|\det(\text{Hess}(\psi(\vec{v}^0)))\right|^{-1/2} \nonumber\\
    &\times& \, e^{i\tfrac{\pi}{4}\text{sign}(\text{Hess}(\psi(\vec{v}_0)))} \left(\frac{2\pi}{\chi}\right)^{2/2} \mathcal{F}(\vec{v}_0)\nonumber\\
    &+& \mathcal{O}(\chi^{-1/2}),\IEEEeqnarraynumspace
    \label{Stationary_aprox}
\end{IEEEeqnarray}
where $\text{Hess}(\psi)$ denotes the Hessian matrix of $\psi$ and $\text{sign}(\text{Hess}(\psi))$ is the number of positive eigenvalues minus the number of negative eigenvalues of the Hessian matrix. For our case, the phase $\psi(x,v_1,v_2)$ has the form
\begin{IEEEeqnarray}{rCl}
    \psi(x,v_1,v_2) & = & \frac{x^3} {3 \omega_q^2}\left(v_1^4 \omega_{p_1}^2+2 v_1^3 \omega_{p_1} ((2 v_2-1) \omega_{p_1}-2 v_2 \omega_q) \right.\nonumber\\
    & & +\: v_1^2 \omega_{p_1} \left((6 v_2^2-6 v_2+1) \omega_{p_1}-6 (v_2-1) v_2 \omega_q\right)\nonumber\\
    & & +\:2 v_1 \left(2 v_2^2-3 v_2+1\right) v_2 \omega_{p_1} (\omega_{p_1}-\omega_q)\nonumber\\
    & &\left. +\:(v_2-1)^2 v_2^2 (\omega_{p_1}-\omega_q)^2\right).
\end{IEEEeqnarray}

\begin{widetext}

\begin{figure}[t]
    \centering
    \includegraphics[width=0.33\textwidth]{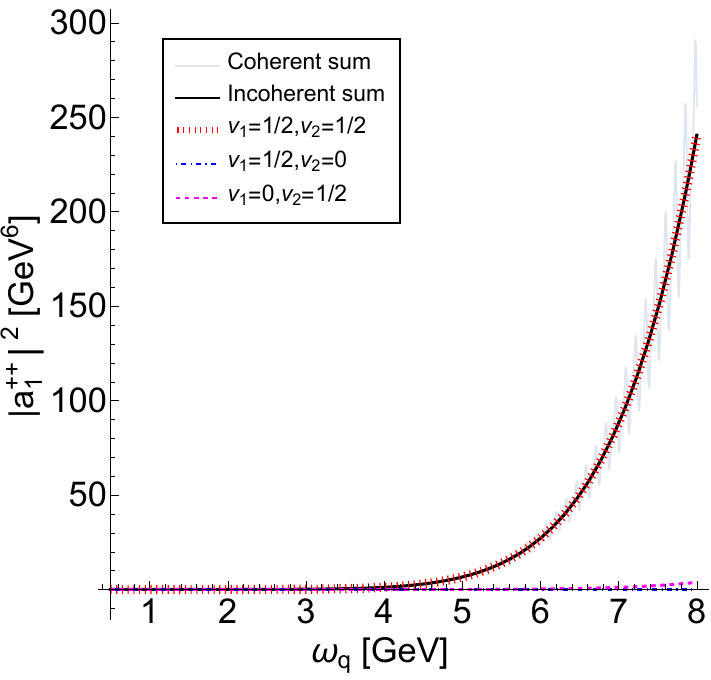}
    \includegraphics[width=0.33\textwidth]{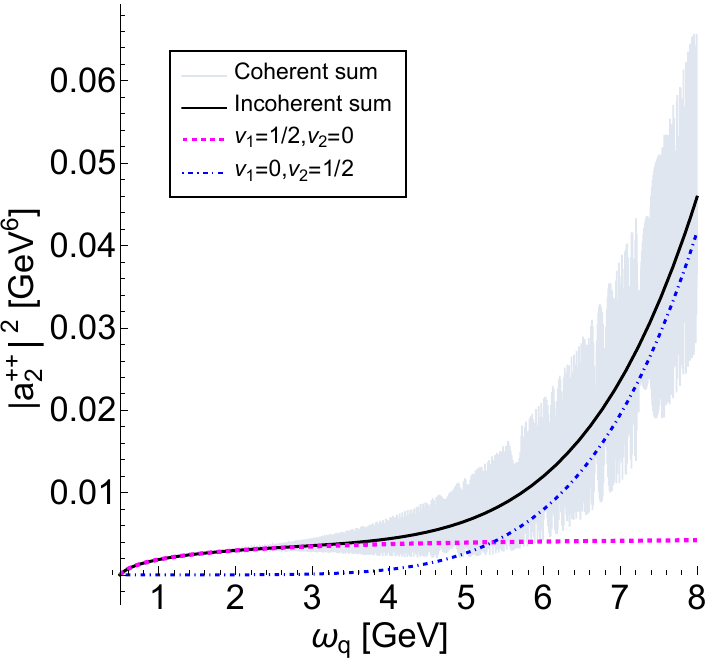}
    \includegraphics[width=0.33\textwidth]{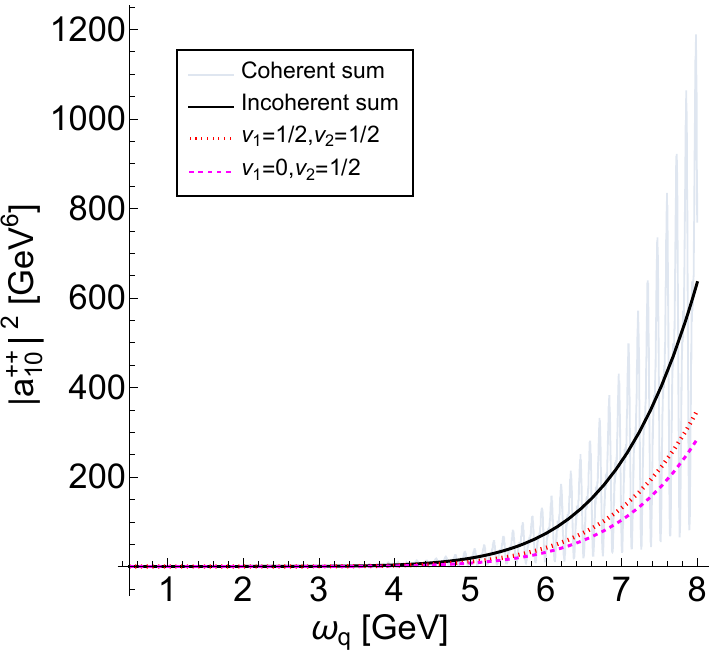}
    \caption{Coefficients $|a_1^{++}|^2$ (left), $|a_2^{++}|^2$ (center) and  $|a_{10}^{++}|^2$ (right) as functions of the photon energy \textcolor{black}{for the case where the angle between the direction of the photon propagation and the magnetic field is} $\theta=\pi/2$. The energy of one of the gluons is fixed to $\omega_{p_1}=0.5$ GeV whereas the energy of the other gluon is given by energy conservation. The magnetic field is set to $|eB|=m_{\pi}^2$. The contribution from the three light quark flavors $u,d,s$, is accounted for. Three maxima in the $(v_1,v_2)$ space (see text) contribute to the coefficient $|a_1^{++}|^2$, whereas two contribute to the coefficients $|a_2^{++}|^2$ and $|a_{10}^{++}|^2$. The shaded showing region, showing rapid oscillations, corresponds to the coherent sum of the contributions from each maxima. For the coefficient $|a_1^{++}|^2$, the contribution from the three contributing maxima are shown whereas for the coefficients $|a_2^{++}|^2$ and $|a_{10}^{++}|^2$ the contributions from the two contributing maxima are shown.  Notice that in each case, the incoherent sum corresponds to the average of the coherent sum.} 
    \label{Gcom}
\end{figure}

We find that the maxima, that is, the points where $\nabla \psi = 0$ in the integrands of $J_i$, $i=1,2,3$, are located at $(v_1,v_2) = (1/2,1/2),\; (1/2,0)$ and $(0,1/2)$. Therefore, in general, the phase is stationary in these three local maxima. We therefore need to consider these contributions for our approximation.

\end{widetext}

\section{One-loop functional behavior of the tensor basis coefficients in the intermediate field strength limit}\label{IV}

We now discuss the behavior of the tensor basis coefficients in the one-loop approximation and the intermediate field strength limit. Figure~\ref{Gcom} shows the behavior of the square of the three basis coefficients as functions of the photon energy for a fixed value of the field strength $|eB|=m_\pi^2$, a fixed energy of one of the gluons $\omega_{p_1}=0.5$ GeV and a fixed angle between the direction of motion of the photon and the direction of the magnetic field, $\theta=\pi/2$.  Hereafter it is to be understood that the energy of the second gluon is fixed by energy-momentum conservation. The figure shows the contribution from the different local maxima when using the stationary phase approximation. The contributing points in the $(v_1,v_2)$ plane are indicated in the insets. Notice that the coherent sum of these contributions, indicated by the shaded region, shows rapid oscillations which are the result of the interference of these terms, since each individual contribution is in general complex. If we instead carry out the incoherent sum of the contributions we obtain the smooth curves shown in the graphs. The incoherent sum coincides with the average of the coherent sum. Therefore, hereafter we show the results for the square of the amplitudes obtained from the incoherent sum of the different terms contributing to the amplitudes.

Figure~\ref{Gw} shows the behavior of the square of the three basis coefficients as functions of the photon energy for a fixed value of the field strength $|eB|=m_\pi^2$, and a fixed energy of one of the gluons $\omega_{p_1}=0.5$ GeV. The figure shows the square of the amplitudes when varying the angle between the direction of propagation of the photon and the direction of the magnetic field. Notice that the maximum of each of the square of the coefficients is obtained for $\theta=\pi/2$. This means that a positive $\nu_2$ can be expected form photons emitted by this kind of processes.

Figure~\ref{Gtheta} shows the behavior of the square of the three basis coefficients as functions of the angle $\theta$ between the propagation of the photon and the direction of the magnetic field for a fixed value of the field strength $|eB|=m_\pi^2$, and a fixed energy of one of the gluons $\omega_{p_1}=0.5$ GeV for different photon energies. Notice that the maximum amplitude squared is obtained for $\theta=\pi/2$ and for the largest photon energy. Since we work in the intermediate field strength regime, $q_\perp^2/|eB|<1$ and $q_\perp^2=\omega_q^2\sin^2\theta$, for a given photon energy, the angle $\theta$ cannot be that close to $\theta\sim 0,2\pi$ and this is implemented with the chosen $\theta$ domain.

Figure~\ref{eB} shows the behavior of the square of the three basis coefficients as functions of the field strength for a fixed photon energy $\omega_q=4$ GeV and a fixed energy for one of the gluons $\omega_{p_1}=0.5$ GeV for different values of the angle  $\theta$ between the direction of motion of the photon and the direction of the magnetic field. Notice that the maximum value of the square of the amplitudes is obtained for the smallest field strength and, as before, for $\theta=\pi/2$.

Figure~\ref{T} shows the behavior of the sum of the squares of the three basis coefficients as functions of the photon energy (left), the angle $\theta$ between the direction of propagation of the photon and the magnetic field (center), and the field strength (right). 

\begin{widetext}

\begin{figure}[t]
    \centering
    \includegraphics[width=0.33\textwidth]{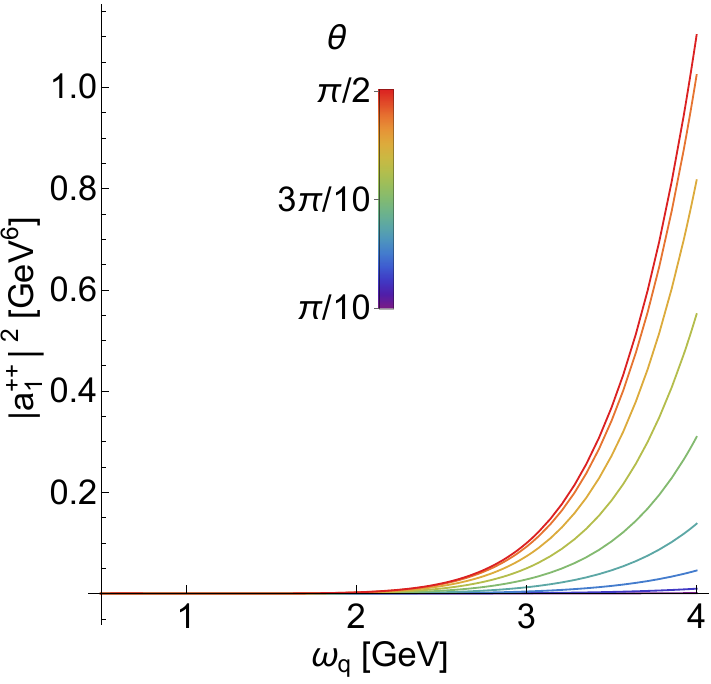}
     \includegraphics[width=0.33\textwidth]{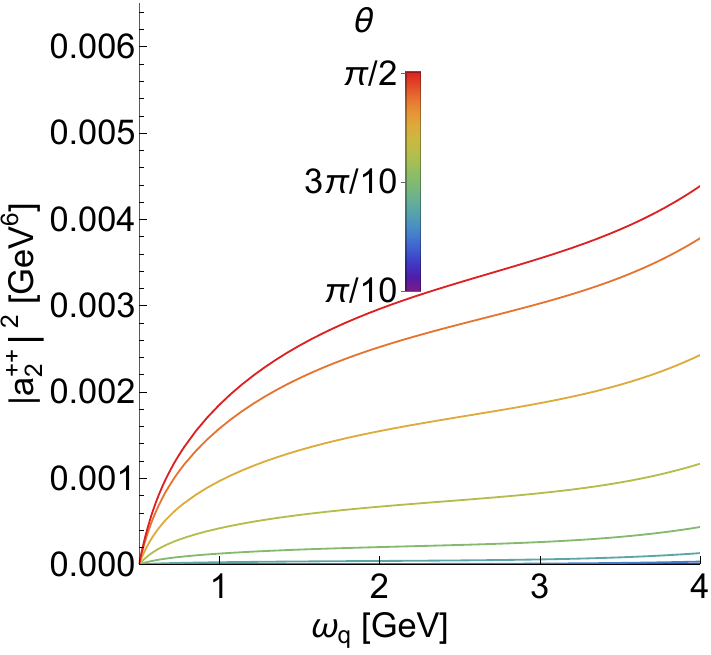}
      \includegraphics[width=0.33\textwidth]{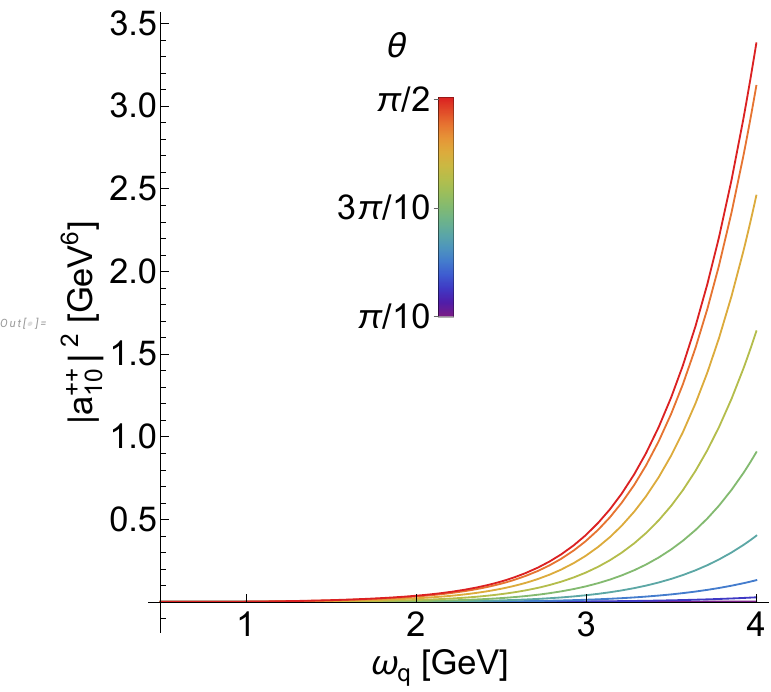}
    \caption{Coefficients $|a_1^{++}|^2$ (left), $|a_2^{++}|^2$ (center) and  $|a_{10}^{++}|^2$ (right) as functions of the photon energy. The energy of one of the gluons is fixed to $\omega_{p_1}=0.5$ GeV, whereas the energy of the other gluon is given by energy conservation. The magnetic field is set to $|eB|=m_{\pi}^2$. The contribution from the three light quark flavors $u,d,s$, is accounted for. Notice that the favored angles for photon emission are close to $\theta=\pi/2.$ } 
    \label{Gw}
\end{figure}

The total amplitude squared shows similar features to those of the individual contributions, namely, favored emission angles close to $\theta=\pi/2$, monotonically increasing coefficients as functions of the photon energy and a mild dependence on the field strength.

\begin{figure}[t]
    \centering
    \includegraphics[width=0.33\textwidth]{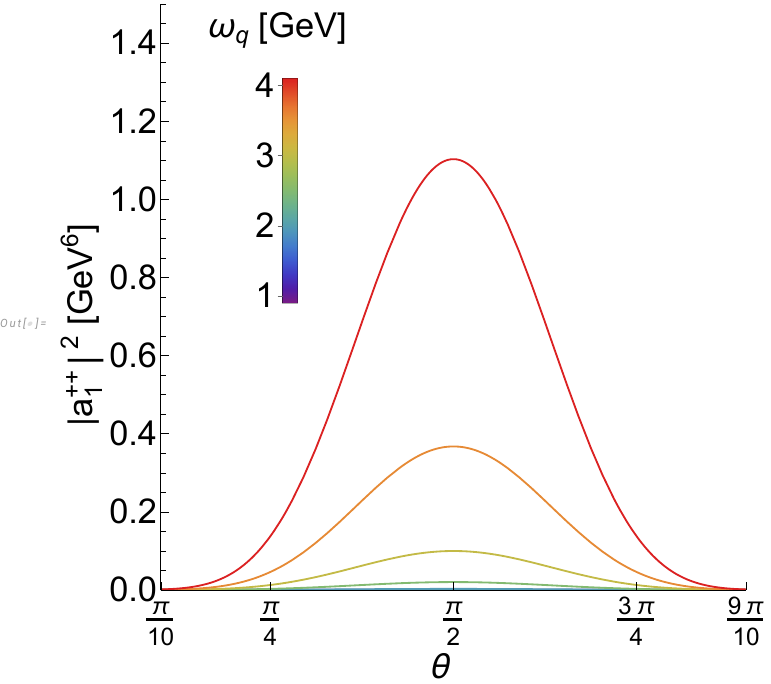}
     \includegraphics[width=0.33\textwidth]{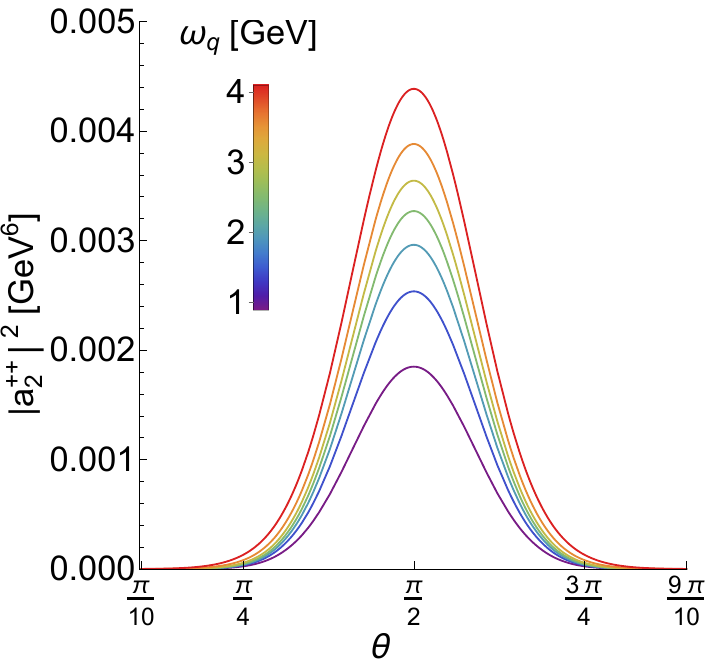}
      \includegraphics[width=0.33\textwidth]{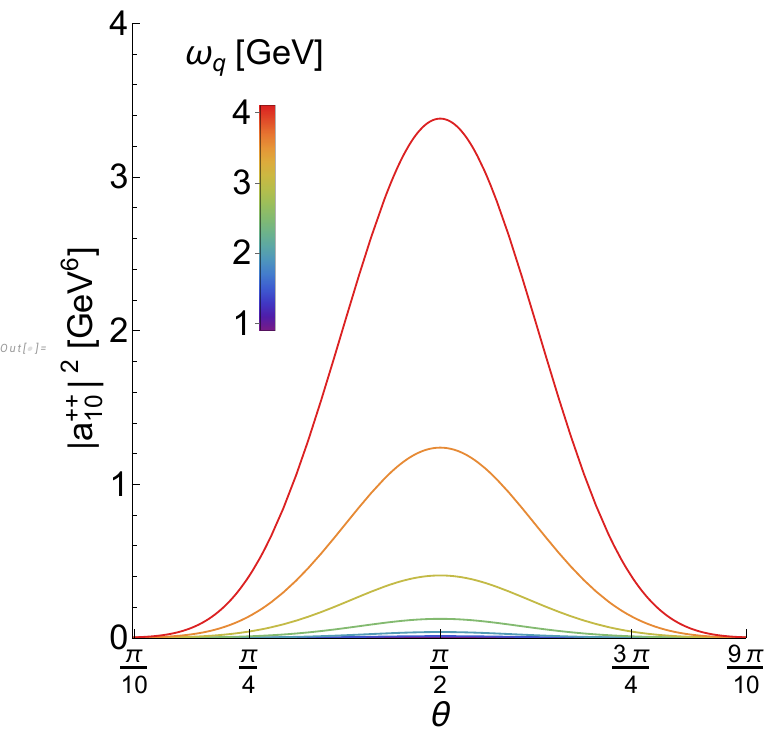}
    \caption{Coefficients $|a_1^{++}|^2$ (left), $|a_2^{++}|^2$ (center) and  $|a_{10}^{++}|^2$ (right) as functions of the angle $\theta$ between the photon momentum and the magnetic field for different values of the photon energy $\omega_q$. The energy of one of the gluons is fixed to $\omega_{p_1}=0.5$ GeV, whereas the energy of the other gluon is given by energy conservation. The magnetic field is set to $|eB|=m_{\pi}^2$. The contribution from the three light quark flavors $u,d,s$, is accounted for. Notice that the favored angles for photon emission are close to $\theta=\pi/2$. Since the approximation requires that $q_\perp^2/|eB|<1$, and $q_\perp^2=\omega_q^2\sin^2\theta$, for a given photon energy, the angle $\theta$ cannot be that close to $\theta\sim 0,2\pi$ and this is implemented with the chosen $\theta$ domain.} 
    \label{Gtheta}
\end{figure}

\begin{figure}[h]
\centering
\includegraphics[width=0.325\textwidth]{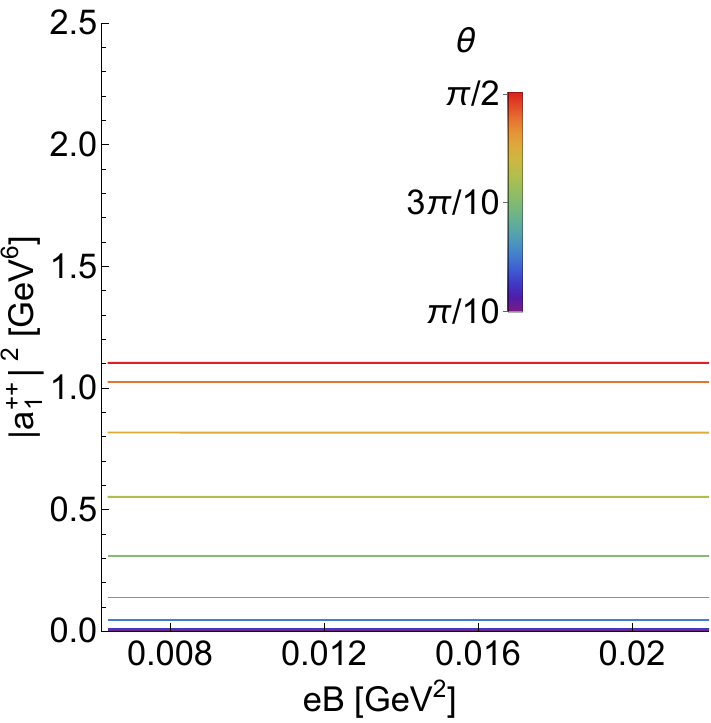}
\includegraphics[width=0.34\textwidth]{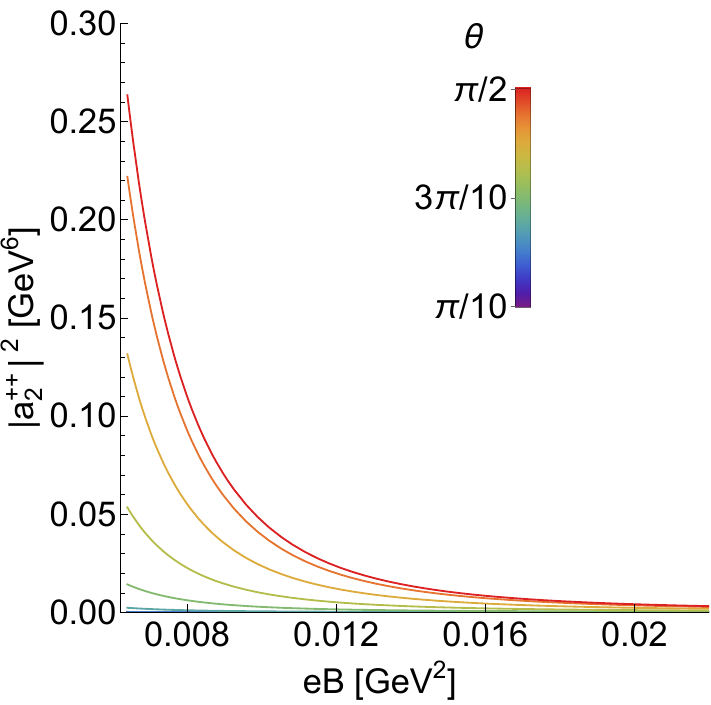}
\includegraphics[width=0.325\textwidth]{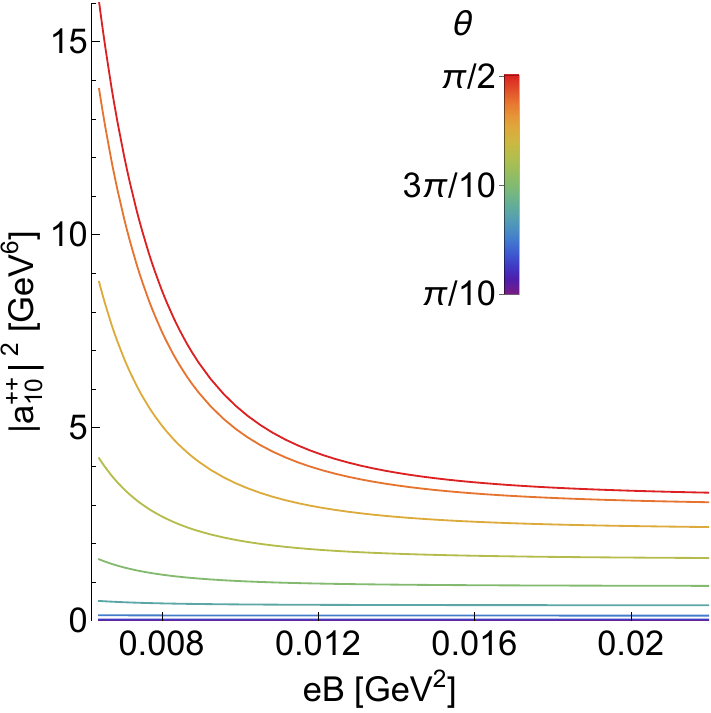}
\caption{Coefficients $|a_1^{++}|^2$ (left), $|a_2^{++}|^2$ (center) and  $|a_{10}^{++}|^2$ (right) as function of the magnetic interaction $eB$. The energies of the photon and one gluon are fixed to $\omega_q=4$ GeV and $\omega_{p_1}=0.5$ GeV, whereas the energy of the other gluon is given by energy conservation. The contribution from the three light quark flavors $u,d,s$, is accounted for. Notice that the favored angles for photon emission are close to $\theta=\pi/2.$} 
    \label{eB}
\end{figure}

\begin{figure}[h]
    \centering
   \includegraphics[width=0.33\textwidth]{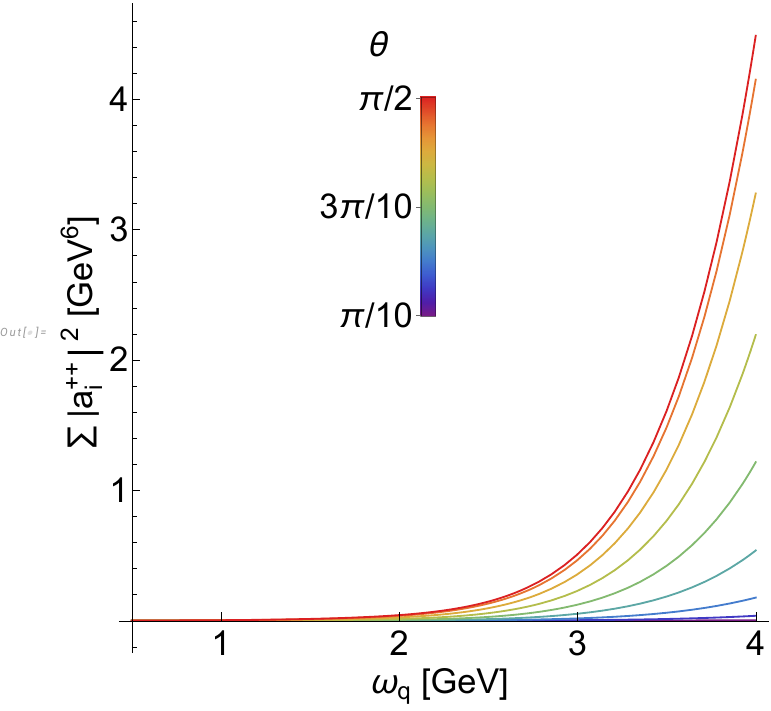}
    \includegraphics[width=0.33\textwidth]{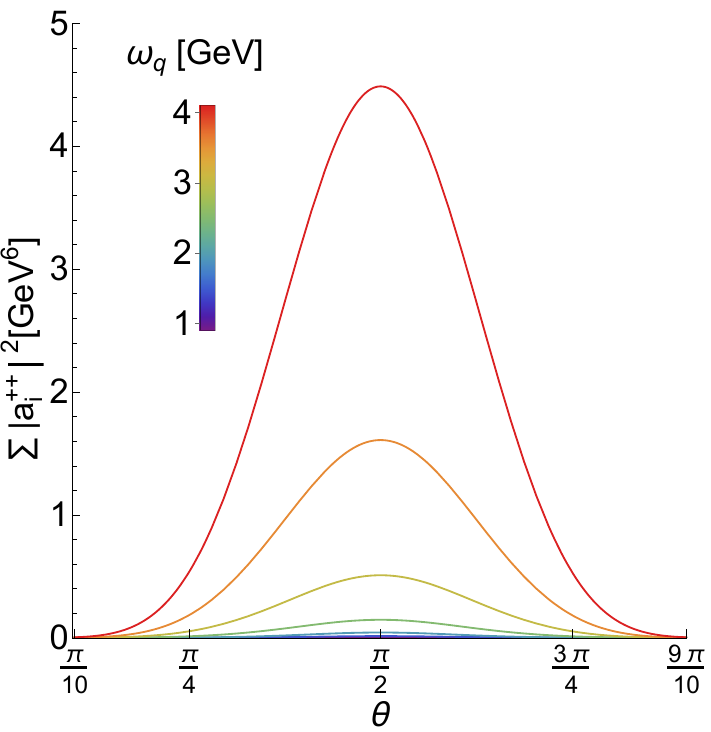}
      \includegraphics[width=0.33\textwidth]{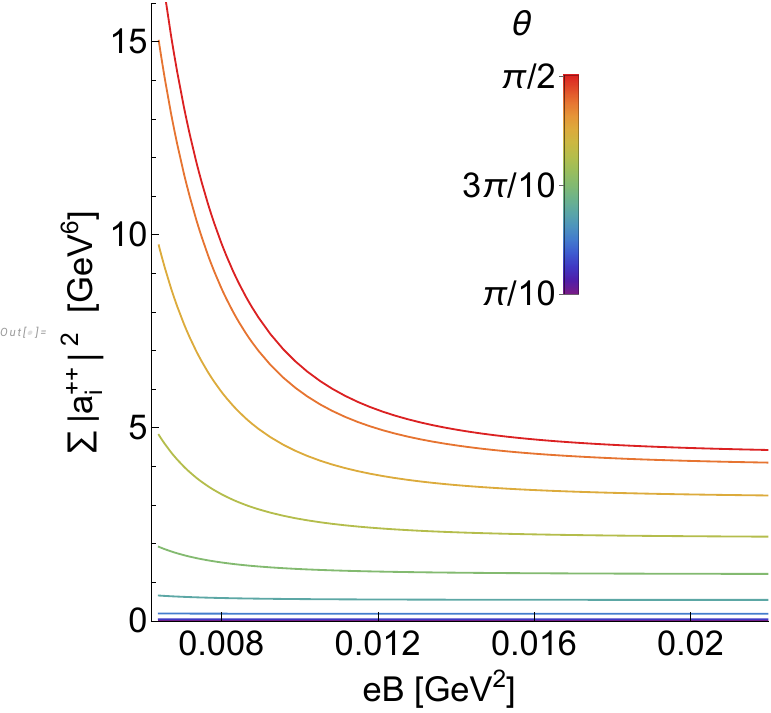}
    \caption{Sum of the squared amplitudes, $|a_1^{++}|^2$, $|a_2^{++}|^2$ and $|a_{10}^{++}|^2$, as functions of the photon energy $\omega_q$ (left), the angle $\theta$ between the photon momentum and the magnetic field (center), and the field strength $|eB|$ (right). In the tree cases, the energy of one of the gluons is fixed to $\omega_{p_1}=0.5$ GeV, whereas the energy of the other gluon is given by energy conservation.} 
    \label{T}
\end{figure}

\section{Summary and Conclusions}\label{concl}

In this work we have found the general structure for the two-gluon one-photon vertex in the presence of a homogeneous and constant magnetic field. We have shown that, after enforcing the symmetries satisfied by the strong and electromagnetic interactions under charge and parity conjugation and gluon exchange, and for gluons and photons on mass-shell, the possible tensor structures describing the vertex reduce to three. These are expressed as external products of the polarization vectors for each of the particles in the vertex. In order to explicitly explore the properties of these coefficients, we have performed a one-loop calculation in the intermediate field regime, which is the most appropriate one to describe the possible effects of the presence of a magnetic field as a catalyzer of photon emission during pre-equilibrium in a heavy-ion reaction. We have shown that the most favored direction of photon propagation is the direction transverse to the magnetic field. This is consistent with a positive contribution to $\nu_2$ which can potentially help to explain the larger than expected photon elliptic flow coefficient measured in relativistic heavy-ion collisions. \textcolor{black}{We are now in a position to use the findings of this work for the computation of the photon yield and elliptic flow coefficient. Following the discussion of Ref.~\cite{Ayala:2017vex}, this entails computing the matrix element squared summed over the possible polarizations, which is equivalent to the sum of the squares of the coefficients of the three polarization tensors. In this manner, the photon yield will be obtained weighing the matrix element squared with the initial/final state gluon distributions at pre-equilibrium for gluon fusion/splitting and integrating over phase space. The $\nu_2$ coefficient will be computed from the weighed average of the Fourier decomposition of the azimuthal distribution, that also involves the computation of the total yield which is in turn obtained from the overall phase space integration of the matrix element squared.} The consequences of this analysis to explain the photon puzzle are currently being explored when accounting for the large occupation number of gluons during pre-equilibrium and the results will be soon reported elsewhere.

\section*{Acknowledgements}

S.B.-L and J.J. M.-S. each acknowledge the financial support of fellowships granted by Consejo Nacional de Humanidades, Ciencia y Tecnolog\'ia as part of the Sistema Nacional de Posgrados. Support for this work was received in part by UNAM-Programa de Apoyo a Proyectos de Investigación  e Innovación Tecnológica (PAPIIT) Grant No. IG100322 and by Consejo Nacional de Humanidades, Ciencia y Tecnolog\'ia Grant No. CF-2023-G-433.

\end{widetext}

\section{Appendix: Elements for the explicit computation of the tensor basis coefficients}
As an example of the general method used to compute the coefficients of the tensor structures, we compute the $a_1^{++}$ coefficient, which is given by
\begin{eqnarray}
a_1^{++}&=&\hat{l}_{(p_1a)_\mu} \hat{l}_{(p_2b)_\nu} \hat{l}_{(q)_\alpha}\left. \Gamma^{\mu\nu\alpha}_{ab}(p_1,p_2,q)\right|_{\mbox{\tiny{on-shell}}}.
\label{firstofeqsapp}
\end{eqnarray}
To compute this coefficient, we need first to find the residues at the poles. There are three contributions, explicitly,
\begin{IEEEeqnarray}{rCl}
\label{Rs1}
R_1(n,v_1,v_2)&=&\text{Res}\left(\csc^2(x) e^{i x^3\text{Arg}} G_1(x,v_1,v_2),n\pi\right)\nonumber\\  
&=& \frac{\partial}{\partial x}\left(e^{i x^3\text{Arg}} G_1(x,v_1,v_2)\right)\vert_{x = n\pi},
     \end{IEEEeqnarray}
    \begin{IEEEeqnarray}{rCl}
\label{Rs2}
R_2(n,v_1,v_2)&=&\text{Res}\left(\csc^3(x) e^{i x^3\text{Arg}} G_2(x,v_1,v_2),n\pi\right)\nonumber\\  
&=& \frac{1}{2}(-1)^n\Big{(}e^{i n^3\pi^3\text{Arg}} G_2(n\pi,v_1,v_2)\nonumber\\
&+&\frac{\partial^2}{\partial x^2}\left(e^{i x^3\text{Arg}} G_2(x,v_1,v_2)\right)\vert_{x = n\pi} \Big{)},
\end{IEEEeqnarray}
\begin{IEEEeqnarray}{rCl}
\label{Rs3}
R_3(n,v_1,v_2)&=&\text{Res}\left(\csc^4(x) e^{i x^3\text{Arg}} G_3(x,v_1,v_2),n\pi\right)\nonumber\\  
&=& \frac{1}{6}\Big{(}4e^{i n^3\pi^3\text{Arg}} G_3(n\pi,v_1,v_2)\nonumber\\
&+&\frac{\partial^3}{\partial x^3}\left(e^{i x^3\text{Arg}} G_3(x,v_1,v_2)\right)\vert_{x = n\pi}  \Big{)}.
\end{IEEEeqnarray}

The stationary phase approximation, referred to in Eq.~(\ref{Stationary_aprox}), is now used to compute each of the terms in Eqs.~(\ref{Rs1}),~(\ref{Rs2}) and~(\ref{Rs3}) to then perform the sum. For this purpose, we take 
$\text{Arg}=\chi x^3\psi(v_1,v_2)$, with $\chi=\frac{q_\perp^2}{|q_f B|}$,
to obtain
\begin{IEEEeqnarray}{rCl}
   \det(\text{Hess}(\psi(1/2,1/2))) &=& -\frac{\omega_{p_1}(\omega_{p_1}-\omega_q)}{3\omega_q^2},
        \end{IEEEeqnarray}
    \begin{IEEEeqnarray}{rCl}
   \det(\text{Hess}(\psi(1/2,0))) &=& \frac{\omega_{p_1}^2(\omega_{p_1}-\omega_q)}{3\omega_q^3},
        \end{IEEEeqnarray}
    \begin{IEEEeqnarray}{rCl}
   \det(\text{Hess}(\psi(0,1/2))) &=& -\frac{\omega_{p_1}(\omega_{p_1}-\omega_q)^2}{3\omega_q^3}.
\end{IEEEeqnarray}
Therefore, the stationary phase approximation for each of the above terms, evaluated at $n\pi$, can be obtained using the expressions
\begin{IEEEeqnarray}{rCl}
     H_1(n,\vec{v})&=&\left(\frac{2\pi}{\chi}\right)\frac{R_1(n,\vec{v})}{\left|\det(\text{Hess}(\psi(\vec{v})))\right|^{1/2}}\nonumber\\
     &\times&e^{i\tfrac{\pi}{4}\text{sign}(\text{Hess}(\psi(\vec{v})))},
     \end{IEEEeqnarray}
    \begin{IEEEeqnarray}{rCl} 
      H_2(n,\vec{v})&=&\left(\frac{2\pi}{\chi}\right)\frac{R_2(n,\vec{v})}{\left|\det(\text{Hess}(\psi(\vec{v})))\right|^{1/2}}\nonumber\\
     &\times&e^{i\tfrac{\pi}{4}\text{sign}(\text{Hess}(\psi(\vec{v})))},
          \end{IEEEeqnarray}
    \begin{IEEEeqnarray}{rCl}
         H_3(n,\vec{v})&=&\left(\frac{2\pi}{\chi}\right)\frac{R_3(n,\vec{v})}{\left|\det(\text{Hess}(\psi(\vec{v})))\right|^{1/2}}\nonumber\\
     &\times&e^{i\tfrac{\pi}{4}\text{sign}(\text{Hess}(\psi(\vec{v})))},
\end{IEEEeqnarray}
evaluated at each of the maxima. Explicitly, these are given by
\begin{widetext}
\begin{IEEEeqnarray}{rCl}
   H_1(n,1/2,1/2)&=&\frac{\sqrt{3} g^2 (-1)^n q_f |q_f B|^2 {\mbox{Tr}}[t_at_b] \csc (\theta ) \left(2 \omega_{p_1}^2-3 \omega_{p_1} \omega_{q}+\omega_{q}^2\right) }{8 \pi ^3 n^2 \omega_{q}^3 \sqrt{\left| \frac{\omega_{p_1} (\omega_{p_1}-\omega_{q})}{\omega_{q}^2}\right| }}e^{\frac{i \pi ^3 n^3 \omega_{q}^2 \sin ^2(\theta )}{48 |q_f B|}}\nonumber\\
   &+&\frac{\sqrt{3} g^2 (-1)^n \left((-1)^n-1\right) q_f {\mbox{Tr}}[t_at_b] \csc (\theta ) \left(2 \omega_{p_1}^2-3 \omega_{p_1} \omega_{q}+\omega_{q}^2\right) }{2048 \pi ^3 n^2 \omega_{q}^3 \sqrt{\left| \frac{\omega_{p_1} (\omega_{p_1}-\omega_{q})}{\omega_{q}^2}\right| }}e^{\frac{i \pi ^3 n^3 \omega_{q}^2 \sin ^2(\theta )}{48 |q_f B|}}\nonumber\\
   &\times&\left(\pi ^4 n^4 \omega_{q}^4 \sin ^4(\theta )-64 i \pi  n |q_f B| \omega_{q}^2 \sin ^2(\theta )\right),
\end{IEEEeqnarray}

\begin{IEEEeqnarray}{rCl}
    H_1(n,1/2,0)&=&\frac{i \sqrt{3} g^2 (-1)^n q_f |q_f B|^2 {\mbox{Tr}}[t_at_b] \omega_{p_1} \csc (\theta ) (\omega_{p_1}-\omega_{q})^2}{4 \pi ^3 n^2 \omega_{q}^4 \sqrt{\left| \frac{\omega_{p_1}^2 (\omega_{p_1}-\omega_{q})}{\omega_{q}^3}\right| }}e^{\frac{i \pi ^3 n^3 \omega_{p_1}^2 \sin ^2(\theta )}{48 |q_f B|}}\nonumber\\
    &+&\frac{\sqrt{3} g^2 (-1)^n \left((-1)^n-1\right) q_f {\mbox{Tr}}[t_at_b] \omega_{p_1} \csc (\theta ) (\omega_{p_1}-\omega_{q}) }{1024 \pi ^4 n^3 \omega_{q}^8 \sqrt{\left| \frac{\omega_{p_1}^2 (\omega_{p_1}-\omega_{q})}{\omega_{q}^3}\right| }}e^{\frac{i \pi ^3 n^3 \omega_{p_1}^2 \sin ^2(\theta )}{48 |q_f B|}}\nonumber\\
    &\times&\Big{(}i \pi ^5 n^5 \omega_{p_1}^4 \omega_{q}^4 \sin ^4(\theta ) (\omega_{p_1}-\omega_{q})+32 i \pi ^3 n^3 \omega_{p_1}^5 \omega_{q}^4 \sin ^4(\theta )\nonumber\\
    &+&64 \pi ^2 n^2 |q_f B| \omega_{p_1}^2 \omega_{q}^4 \sin ^2(\theta ) (\omega_{p_1}-\omega_{q})+512 |q_f B| \omega_{p_1}^3 \omega_{q}^4 \sin ^2(\theta )\Big{)},
\end{IEEEeqnarray}
\begin{IEEEeqnarray}{rCl}
  H_1(n,0,1/2)&=& -\frac{i \sqrt{3} g^2 (-1)^n q_f |q_f B|^2 {\mbox{Tr}}[t_at_b] \csc (\theta ) (\omega_{p_1}-\omega_{q})^3 }{4 \pi ^3 n^2 \omega_{q}^4 \sqrt{\left| \frac{\omega_{p_1} (\omega_{p_1}-\omega_{q})^2}{\omega_{q}^3}\right| }}e^{\frac{i \pi ^3 n^3 \sin ^2(\theta ) (\omega_{p_1}-\omega_{q})^2}{48 |q_f B|}}\nonumber\\
  &-&\frac{i \sqrt{3} g^2 (-1)^n \left((-1)^n-1\right) q_f {\mbox{Tr}}[t_at_b] \csc (\theta ) (\omega_{p_1}-\omega_{q})^3 }{1024 \pi ^3 n^2 \omega_{q}^8 \sqrt{\left| \frac{\omega_{p_1} (\omega_{p_1}-\omega_{q})^2}{\omega_{q}^3}\right| }}e^{\frac{i \pi ^3 n^3 \sin ^2(\theta ) (\omega_{p_1}-\omega_{q})^2}{48 |q_f B|}}\nonumber\\
  &\times&\left(\pi ^4 n^4 \omega_{q}^4 \sin ^4(\theta ) (\omega_{p_1}-\omega_{q})^4-64 i \pi  n |q_f B| \omega_{q}^4 \sin ^2(\theta ) (\omega_{p_1}-\omega_{q})^2\right),
\end{IEEEeqnarray}
\begin{IEEEeqnarray}{rCl}
   H_2(n,1/2,1/2) &=& \frac{-\sqrt{3} g^2 q_f {\mbox{Tr}}[t_at_b] \sin (\theta ) (\omega_{p_1}-\omega_{q})^2 (\omega_{q}-2 \omega_{p_1}) }{32 \pi ^4 (2 n-1)^3 \omega_{q}^2 \sqrt{\left| \frac{\omega_{p_1} (\omega_{p_1}-\omega_{q})}{\omega_{q}^2}\right| }}e^{\frac{i \pi ^3 (2 n-1)^3 \omega_{q}^2 \sin ^2(\theta )}{48 |q_f B|}}\nonumber\\
   &\times&\left(\pi ^3 (2 n-1)^3 \omega_{q}^2 \sin ^2(\theta )-16 i |q_f B|\right),
\end{IEEEeqnarray}

\begin{IEEEeqnarray}{rCl}
     H_2(n,1/2,0)&=&\frac{\sqrt{3} g^2 q_f {\mbox{Tr}}[t_at_b] \omega_{p_1}^3 \sin (\theta ) (\omega_{p_1}-\omega_{q}) }{32 \pi ^4 (2 n-1)^3 \omega_{q}^4 \sqrt{\left| \frac{\omega_{p_1}^2 (\omega_{p_1}-\omega_{q})}{\omega_{q}^3}\right| }}e^{\frac{i \pi ^3 (2 n-1)^3 \omega_{p_1}^2 \sin ^2(\theta )}{48 |q_f B|}}\nonumber\\
     &\times&\left(-i \pi ^3 (2 n-1)^3 \omega_{p_1}^3 \sin ^2(\theta )-2 \left(3 \pi ^2 (1-2 n)^2+8\right) |q_f B| \omega_{p_1}+6 \pi ^2 (1-2 n)^2 |q_f B| \omega_{q}\right),
\end{IEEEeqnarray}

\begin{IEEEeqnarray}{rCl}
     H_2(n,0,1/2)&=& \frac{\sqrt{3} g^2 q_f {\mbox{Tr}}[t_at_b] \sin (\theta ) (\omega_{p_1}-\omega_{q})^3 }{32 \pi ^4 (2 n-1)^3 \omega_{q}^4 \sqrt{\left| \frac{\omega_{p_1} (\omega_{p_1}-\omega_{q})^2}{\omega_{q}^3}\right| }}e^{\frac{i \pi ^3 (2 n-1)^3 \sin ^2(\theta ) (\omega_{p_1}-\omega_{q})^2}{48 |q_f B|}}\nonumber\\
     &\times&\Big{(}2 |q_f B| \Big{(}\left(3 \pi ^2 (1-2 n)^2+8\right) \omega_{p_1}^2+2 \left(4-3 \pi ^2 (1-2 n)^2\right) \omega_{p_1} \omega_{q}+3 \pi ^2 (1-2 n)^2 \omega_{q}^2\Big{)}\nonumber\\
     &+&i \pi ^3 (2 n-1)^3 \omega_{p_1} \sin ^2(\theta ) (\omega_{p_1}-\omega_{q})^2 (\omega_{p_1}+\omega_{q})\Big{)},
\end{IEEEeqnarray}

\begin{IEEEeqnarray}{rCl}
      H_3(n,1/2,0)&=&\frac{-i g^2 q_f {\mbox{Tr}}[t_at_b]\omega_{p_1}^4 \sin\theta (\omega_{p_1}-\omega_q)}{4096 \sqrt{3} \pi ^4 (2 n-1)^3 |q_f B|^2 \omega_q^4 \sqrt{\left| \frac{\omega_{p_1}^2 (\omega_{p_1}-\omega_q)}{\omega_q^3}\right| }} (-1)^{\frac{\pi ^2 (2 n-1)^3 \omega_{p_1}^2 \sin^2\theta}{48 q_f B}} \nonumber\\
    &\times&\left(\pi ^7 (2 n-1)^7 \omega_{p_1}^6 \sin^6\theta+512 \pi  \left(\pi ^2 (1-2 n)^2-4\right) (2 n-1) |q_f B|^2 \omega_{p_1}^2 \sin^2\theta\right.\nonumber\\
    &-&\left. 144 i \pi ^4 (1-2 n)^4 |q_f B| \omega_{p_1}^4 \sin^4\theta-8192 i |q_f B|^3\right).
\end{IEEEeqnarray}
Here, $R_3(n,1/2,1/2)=R_3(n,1/2,0)=0$. Finally, the expression for the $a_1^{++}$ coefficient is
\begin{IEEEeqnarray}{rCl}
   a_1^{++} & \approx & \sum_{\vec{v}^0\in \Upsilon}\sum_n^\infty \Big{(} H_1(n,\vec{v_0})+H_2(n,\vec{v_0})+H_3(n,\vec{v_0}) \Big{)}
   + \sum_{i=1}^3 \int_0 ^\infty d\tau \csc^{j+1} (\tau e^{i\pi / 6}) 
    e^{-\tfrac{q_\perp ^2}{q_f B}\tau^3 F(v_1,v_2)} G_j(\tau e^{i\pi / 6}, v_1, v_2).
\end{IEEEeqnarray}
We can evaluate the previous sums using sum regularization schemes such as Ramanujan or Dirichlet. After implementing the regularization, we obtain

\begin{IEEEeqnarray}{rCl}
    \sum_{n=1}^\infty H_1(n,1/2,1/2) &=& \frac{\sqrt{3} g^2 q_f {\mbox{Tr}}[t_at_b] \sin\theta (\omega_{p_1}-\omega_q)^2 (\omega_q-2 \omega_{p_1}) (-1)^{\frac{\pi ^2 \omega_q^2 \sin^2\theta)}{48 q_f B}} \left(\pi ^3 \omega_q^2 \sin^2\theta+28 i |q_f B| \zeta (3)\right)}{64 \pi ^4 \omega_q^2 \sqrt{\left| \frac{\omega_{p_1} (\omega_{p_1}-\omega_q)}{|\omega_q^2}\right| }},
\end{IEEEeqnarray}    

\begin{IEEEeqnarray}{rCl}
    \sum_{n=1}^\infty H_1(n,1/2,0) &=&
\frac{-g^2 q_f {\mbox{Tr}}[t_at_b]\csc\theta\; (\omega_{p_1}-\omega_q) (2 \omega_{p_1}-\omega_q)  }{2048 \sqrt{3} \pi ^3 \omega_q^3 \sqrt{\left| \frac{\omega_{p_1} (\omega_{p_1}-\omega_q)}{\omega_q^2}\right| }}\Big{(}-768 |q_f B|^2 \text{Li}_2\left(-(-1)^{\frac{\pi ^2 \omega_q^2 \sin ^2 \theta}{48 |q_f B|}}\right)\nonumber\\
&+&\pi  \omega_q^2 \sin^2\theta (-1)^{\frac{\pi ^2 \omega_q^2 \sin^2\theta}{48 |q_f B|}} \left(\pi ^3 \omega_q^2 \sin^2\theta+i\;192 \gamma_E  |q_f B|\right)\Big{)},
\end{IEEEeqnarray}

\begin{IEEEeqnarray}{rCl}
    \sum_{n=1}^\infty H_1(n,0,1/2)&=&\frac{\sqrt{3} g^2 q_f {\mbox{Tr}}[t_at_b]\sin\theta\; (\omega_{p_1}-\omega_q)^3}{128 \pi ^4 \omega_q^4 \sqrt{\left| \frac{\omega_{p_1} (\omega_{p_1}-\omega_q)^2}{\omega_q^3}\right| }} (-1)^{\frac{\pi ^2 \sin^2\theta (\omega_{p_1}-\omega_q)^2}{48 |q_f B|}} \Big{(}i \pi ^3 \omega_{p_1} \cos(2 \theta )(\omega_{p_1}+\omega_q) (\omega_{p_1}-\omega_q)^2\nonumber\\
    &+&\:12 \gamma_E  \pi ^2 |q_f B| (\omega_{p_1}-\omega_q)^2-i \omega_{p_1} (\omega_{p_1}+\omega_q) \left(\pi ^3 (\omega_{p_1}-\omega_q)^2+56 i |q_f B| \zeta (3)\right)\Big{)},
\end{IEEEeqnarray}

\begin{IEEEeqnarray}{rCl}
    \sum_{n=1}^\infty H_2(n,1/2,1/2)&=&\frac{-g^2 q_f {\mbox{Tr}}[t_at_b] \csc\theta\; (\omega_{p_1}-\omega_q) (2 \omega_{p_1}-\omega_q)}{2048 \sqrt{3} \pi ^3 \omega_q^3 \sqrt{\left| \frac{\omega_{p_1} (\omega_{p_1}-\omega_q)}{\omega_q^2}\right| }} \Big{(}-768 |q_f B|^2 \text{Li}_2\left(-(-1)^{\frac{\pi ^2 \omega_q^2 \sin ^2 \theta}{48 q_f B}}\right)\nonumber\\
    &+&\pi  \omega_q^2 \sin^2\theta (-1)^{\frac{\pi ^2 \omega_q^2 \sin^2\theta}{48 |q_f B|}} \left(\pi ^3 \omega_q^2 \sin^2\theta+192 \gamma_E  i |q_f B|\right)\Big{)},
\end{IEEEeqnarray}

\begin{IEEEeqnarray}{rCl}
    \sum_{n=1}^\infty H_2(n,1/2,0)&=&\frac{g^2 q_f {\mbox{Tr}}[t_at_b] \sin ^3\theta\;(\omega_{p_1}-\omega_q)}{1024 \sqrt{3} \pi ^4 \omega_q^4 \sqrt{\left| \frac{\omega_{p_1}^2 (\omega_{p_1}-\omega_q)}{\omega_q^3}\right| }}  \Big{(}768 i \pi  |q_f B|^2 \csc^4\theta (\omega_{p_1}-\omega_q) \text{Li}_2\left(-(-1)^{\frac{\pi ^2 \omega_{p_1}^2 \sin ^2 \theta}{48 |q_f B|}}\right)\nonumber\\
    &+&\:\omega_{p_1}^3 \csc^2\theta (-1)^{\frac{\pi ^2 \omega_{p_1}^2 \sin^2\theta}{48 |q_f B|}} \Big{(}192 |q_f B| \left(\gamma_E  \pi ^2 (\omega_{p_1}-\omega_q)+14 \omega_{p_1} \zeta (3)\right)\nonumber\\
    &-&\:i \pi ^3 \omega_{p_1}^2 \sin^2\theta \left(\left(96+\pi ^2\right) \omega_{p_1}-\pi ^2 \omega_q\right)\Big{)}\Big{)},
\end{IEEEeqnarray}

\begin{IEEEeqnarray}{rCl}
    \sum_{n=1}^\infty H_2(n,0,1/2)&=&\frac{-i g^2 q_f {\mbox{Tr}}[t_at_b] \sin\theta \;(\omega_{p_1}-\omega_q)^3}{2048 \sqrt{3} \pi ^3 \omega_q^4 \sqrt{\left| \frac{\omega_{p_1} (\omega_{p_1}-\omega_q)^2}{\omega_q^3}\right| }}  \Big{(}1536 |q_f B|^2 \csc^2\theta \text{Li}_2\left(-(-1)^{\frac{\pi ^2 (\omega_{p_1}-\omega_q)^2 \sin ^2 \theta}{48 |q_f B|}}\right)\nonumber\\
    &+&\:\pi  (\omega_{p_1}-\omega_q)^2 \left(\pi ^3 \cos(2 \theta ) (\omega_{p_1}-\omega_q)^2-384 i \gamma_E  |q_f B|-\pi ^3 (\omega_{p_1}-\omega_q)^2\right)\Big{)}.
\end{IEEEeqnarray}

\begin{IEEEeqnarray}{rCl}
      \sum_{n=1}^\infty H_3(n,1/2,0)&=&\frac{i g^2 q_f {\mbox{Tr}}[t_at_b] \omega_{p_1}^4 \sin\theta (\omega_{p_1}-\omega_q)}{4096 \sqrt{3} \pi ^4 q_f^2 B^2 \omega_q^4 \sqrt{\left| \frac{\omega_{p_1}^2 (\omega_{p_1}-\omega_q)}{\omega_q^3}\right| }} (-1)^{\frac{\pi ^2 (2 n-1)^3 \omega_{p_1}^2 \sin^2\theta}{48 ||q_f B|}} \nonumber\\
    &\times&\left(\pi^7\omega_{p_1}^6 \sin^6(\theta)/10+512 \pi^3 |q_f B|^2 \omega_{p_1}^2 \sin^2\theta + 48 i \pi^4  |q_f B| \omega_{p_1}^4 \sin^4\theta + 7168 i |q_f B|^3 \zeta(3)\right).
\end{IEEEeqnarray}
where $\gamma_E$ is the Euler-Mascheroni constant and $\zeta$ the Riemann zeta function. The other two coefficients are computed in a similar fashion.

\end{widetext}


\bibliography{biblio}

\providecommand{\noopsort}[1]{}\providecommand{\singleletter}[1]{#1}%
\begin{thebibliography}{66}%
\makeatletter
\providecommand \@ifxundefined [1]{%
 \@ifx{#1\undefined}
}%
\providecommand \@ifnum [1]{%
 \ifnum #1\expandafter \@firstoftwo
 \else \expandafter \@secondoftwo
 \fi
}%
\providecommand \@ifx [1]{%
 \ifx #1\expandafter \@firstoftwo
 \else \expandafter \@secondoftwo
 \fi
}%
\providecommand \natexlab [1]{#1}%
\providecommand \enquote  [1]{``#1''}%
\providecommand \bibnamefont  [1]{#1}%
\providecommand \bibfnamefont [1]{#1}%
\providecommand \citenamefont [1]{#1}%
\providecommand \href@noop [0]{\@secondoftwo}%
\providecommand \href [0]{\begingroup \@sanitize@url \@href}%
\providecommand \@href[1]{\@@startlink{#1}\@@href}%
\providecommand \@@href[1]{\endgroup#1\@@endlink}%
\providecommand \@sanitize@url [0]{\catcode `\\12\catcode `\$12\catcode
  `\&12\catcode `\#12\catcode `\^12\catcode `\_12\catcode `\%12\relax}%
\providecommand \@@startlink[1]{}%
\providecommand \@@endlink[0]{}%
\providecommand \url  [0]{\begingroup\@sanitize@url \@url }%
\providecommand \@url [1]{\endgroup\@href {#1}{\urlprefix }}%
\providecommand \urlprefix  [0]{URL }%
\providecommand \Eprint [0]{\href }%
\providecommand \doibase [0]{http://dx.doi.org/}%
\providecommand \selectlanguage [0]{\@gobble}%
\providecommand \bibinfo  [0]{\@secondoftwo}%
\providecommand \bibfield  [0]{\@secondoftwo}%
\providecommand \translation [1]{[#1]}%
\providecommand \BibitemOpen [0]{}%
\providecommand \bibitemStop [0]{}%
\providecommand \bibitemNoStop [0]{.\EOS\space}%
\providecommand \EOS [0]{\spacefactor3000\relax}%
\providecommand \BibitemShut  [1]{\csname bibitem#1\endcsname}%
\let\auto@bib@innerbib\@empty
\bibitem [{\citenamefont {Skokov}\ \emph {et~al.}(2009)\citenamefont {Skokov},
  \citenamefont {Illarionov},\ and\ \citenamefont {Toneev}}]{Skokov:2009qp}%
  \BibitemOpen
  \bibfield  {author} {\bibinfo {author} {\bibfnamefont {V.}~\bibnamefont
  {Skokov}}, \bibinfo {author} {\bibfnamefont {A.~Y.}\ \bibnamefont
  {Illarionov}}, \ and\ \bibinfo {author} {\bibfnamefont {V.}~\bibnamefont
  {Toneev}},\ }\href {\doibase 10.1142/S0217751X09047570} {\bibfield  {journal}
  {\bibinfo  {journal} {Int. J. Mod. Phys. A}\ }\textbf {\bibinfo {volume}
  {24}},\ \bibinfo {pages} {5925} (\bibinfo {year} {2009})},\ \Eprint
  {http://arxiv.org/abs/0907.1396} {arXiv:0907.1396 [nucl-th]} \BibitemShut
  {NoStop}%
\bibitem [{\citenamefont {Voronyuk}\ \emph {et~al.}(2011)\citenamefont
  {Voronyuk}, \citenamefont {Toneev}, \citenamefont {Cassing}, \citenamefont
  {Bratkovskaya}, \citenamefont {Konchakovski},\ and\ \citenamefont
  {Voloshin}}]{Voronyuk:2011jd}%
  \BibitemOpen
  \bibfield  {author} {\bibinfo {author} {\bibfnamefont {V.}~\bibnamefont
  {Voronyuk}}, \bibinfo {author} {\bibfnamefont {V.~D.}\ \bibnamefont
  {Toneev}}, \bibinfo {author} {\bibfnamefont {W.}~\bibnamefont {Cassing}},
  \bibinfo {author} {\bibfnamefont {E.~L.}\ \bibnamefont {Bratkovskaya}},
  \bibinfo {author} {\bibfnamefont {V.~P.}\ \bibnamefont {Konchakovski}}, \
  and\ \bibinfo {author} {\bibfnamefont {S.~A.}\ \bibnamefont {Voloshin}},\
  }\href {\doibase 10.1103/PhysRevC.83.054911} {\bibfield  {journal} {\bibinfo
  {journal} {Phys. Rev. C}\ }\textbf {\bibinfo {volume} {83}},\ \bibinfo
  {pages} {054911} (\bibinfo {year} {2011})},\ \Eprint
  {http://arxiv.org/abs/1103.4239} {arXiv:1103.4239 [nucl-th]} \BibitemShut
  {NoStop}%
\bibitem [{\citenamefont {McLerran}\ and\ \citenamefont
  {Skokov}(2014)}]{McLerran:2013hla}%
  \BibitemOpen
  \bibfield  {author} {\bibinfo {author} {\bibfnamefont {L.}~\bibnamefont
  {McLerran}}\ and\ \bibinfo {author} {\bibfnamefont {V.}~\bibnamefont
  {Skokov}},\ }\href {\doibase 10.1016/j.nuclphysa.2014.05.008} {\bibfield
  {journal} {\bibinfo  {journal} {Nucl. Phys. A}\ }\textbf {\bibinfo {volume}
  {929}},\ \bibinfo {pages} {184} (\bibinfo {year} {2014})},\ \Eprint
  {http://arxiv.org/abs/1305.0774} {arXiv:1305.0774 [hep-ph]} \BibitemShut
  {NoStop}%
\bibitem [{\citenamefont {Bzdak}\ and\ \citenamefont
  {Skokov}(2012)}]{Bzdak:2011yy}%
  \BibitemOpen
  \bibfield  {author} {\bibinfo {author} {\bibfnamefont {A.}~\bibnamefont
  {Bzdak}}\ and\ \bibinfo {author} {\bibfnamefont {V.}~\bibnamefont {Skokov}},\
  }\href {\doibase 10.1016/j.physletb.2012.02.065} {\bibfield  {journal}
  {\bibinfo  {journal} {Phys. Lett. B}\ }\textbf {\bibinfo {volume} {710}},\
  \bibinfo {pages} {171} (\bibinfo {year} {2012})},\ \Eprint
  {http://arxiv.org/abs/1111.1949} {arXiv:1111.1949 [hep-ph]} \BibitemShut
  {NoStop}%
\bibitem [{\citenamefont {Zhang}\ \emph {et~al.}(2023)\citenamefont {Zhang},
  \citenamefont {Feng}, \citenamefont {Ren}, \citenamefont {Hua},\ and\
  \citenamefont {Huo}}]{Zhang:2023ppo}%
  \BibitemOpen
  \bibfield  {author} {\bibinfo {author} {\bibfnamefont {C.-H.}\ \bibnamefont
  {Zhang}}, \bibinfo {author} {\bibfnamefont {Q.-C.}\ \bibnamefont {Feng}},
  \bibinfo {author} {\bibfnamefont {Y.-Y.}\ \bibnamefont {Ren}}, \bibinfo
  {author} {\bibfnamefont {L.-M.}\ \bibnamefont {Hua}}, \ and\ \bibinfo
  {author} {\bibfnamefont {L.}~\bibnamefont {Huo}},\ }\href {\doibase
  10.1038/s41598-023-48705-1} {\bibfield  {journal} {\bibinfo  {journal} {Sci.
  Rep.}\ }\textbf {\bibinfo {volume} {13}},\ \bibinfo {pages} {21500} (\bibinfo
  {year} {2023})}\BibitemShut {NoStop}%
\bibitem [{\citenamefont {Sun}\ and\ \citenamefont
  {Yan}(2023{\natexlab{a}})}]{Sun:2023rhh}%
  \BibitemOpen
  \bibfield  {author} {\bibinfo {author} {\bibfnamefont {J.-A.}\ \bibnamefont
  {Sun}}\ and\ \bibinfo {author} {\bibfnamefont {L.}~\bibnamefont {Yan}},\
  }\href@noop {} {\  (\bibinfo {year} {2023}{\natexlab{a}})},\ \Eprint
  {http://arxiv.org/abs/2311.03929} {arXiv:2311.03929 [nucl-th]} \BibitemShut
  {NoStop}%
\bibitem [{\citenamefont {Adam}\ \emph {et~al.}(2021)\citenamefont {Adam} \emph
  {et~al.}}]{STAR:2019wlg}%
  \BibitemOpen
  \bibfield  {author} {\bibinfo {author} {\bibfnamefont {J.}~\bibnamefont
  {Adam}} \emph {et~al.} (\bibinfo {collaboration} {STAR}),\ }\href {\doibase
  10.1103/PhysRevLett.127.052302} {\bibfield  {journal} {\bibinfo  {journal}
  {Phys. Rev. Lett.}\ }\textbf {\bibinfo {volume} {127}},\ \bibinfo {pages}
  {052302} (\bibinfo {year} {2021})},\ \Eprint
  {http://arxiv.org/abs/1910.12400} {arXiv:1910.12400 [nucl-ex]} \BibitemShut
  {NoStop}%
\bibitem [{\citenamefont {Brandenburg}\ \emph {et~al.}(2021)\citenamefont
  {Brandenburg}, \citenamefont {Zha},\ and\ \citenamefont
  {Xu}}]{Brandenburg:2021lnj}%
  \BibitemOpen
  \bibfield  {author} {\bibinfo {author} {\bibfnamefont {J.~D.}\ \bibnamefont
  {Brandenburg}}, \bibinfo {author} {\bibfnamefont {W.}~\bibnamefont {Zha}}, \
  and\ \bibinfo {author} {\bibfnamefont {Z.}~\bibnamefont {Xu}},\ }\href
  {\doibase 10.1140/epja/s10050-021-00595-5} {\bibfield  {journal} {\bibinfo
  {journal} {Eur. Phys. J. A}\ }\textbf {\bibinfo {volume} {57}},\ \bibinfo
  {pages} {299} (\bibinfo {year} {2021})},\ \Eprint
  {http://arxiv.org/abs/2103.16623} {arXiv:2103.16623 [hep-ph]} \BibitemShut
  {NoStop}%
\bibitem [{\citenamefont {Casta\~no Yepes}\ \emph
  {et~al.}(2023{\natexlab{a}})\citenamefont {Casta\~no Yepes}, \citenamefont
  {Loewe}, \citenamefont {Mu\~noz},\ and\ \citenamefont
  {Rojas}}]{Castano-Yepes:2023brq}%
  \BibitemOpen
  \bibfield  {author} {\bibinfo {author} {\bibfnamefont {J.~D.}\ \bibnamefont
  {Casta\~no Yepes}}, \bibinfo {author} {\bibfnamefont {M.}~\bibnamefont
  {Loewe}}, \bibinfo {author} {\bibfnamefont {E.}~\bibnamefont {Mu\~noz}}, \
  and\ \bibinfo {author} {\bibfnamefont {J.~C.}\ \bibnamefont {Rojas}},\ }\href
  {\doibase 10.1103/PhysRevD.108.116013} {\bibfield  {journal} {\bibinfo
  {journal} {Phys. Rev. D}\ }\textbf {\bibinfo {volume} {108}},\ \bibinfo
  {pages} {116013} (\bibinfo {year} {2023}{\natexlab{a}})},\ \Eprint
  {http://arxiv.org/abs/2308.12249} {arXiv:2308.12249 [hep-th]} \BibitemShut
  {NoStop}%
\bibitem [{\citenamefont {Casta\~no Yepes}\ \emph
  {et~al.}(2023{\natexlab{b}})\citenamefont {Casta\~no Yepes}, \citenamefont
  {Loewe}, \citenamefont {Mu\~noz}, \citenamefont {Rojas},\ and\ \citenamefont
  {Zamora}}]{Castano-Yepes:2022luw}%
  \BibitemOpen
  \bibfield  {author} {\bibinfo {author} {\bibfnamefont {J.~D.}\ \bibnamefont
  {Casta\~no Yepes}}, \bibinfo {author} {\bibfnamefont {M.}~\bibnamefont
  {Loewe}}, \bibinfo {author} {\bibfnamefont {E.}~\bibnamefont {Mu\~noz}},
  \bibinfo {author} {\bibfnamefont {J.~C.}\ \bibnamefont {Rojas}}, \ and\
  \bibinfo {author} {\bibfnamefont {R.}~\bibnamefont {Zamora}},\ }\href
  {\doibase 10.1103/PhysRevD.107.096014} {\bibfield  {journal} {\bibinfo
  {journal} {Phys. Rev. D}\ }\textbf {\bibinfo {volume} {107}},\ \bibinfo
  {pages} {096014} (\bibinfo {year} {2023}{\natexlab{b}})},\ \Eprint
  {http://arxiv.org/abs/2211.16985} {arXiv:2211.16985 [hep-th]} \BibitemShut
  {NoStop}%
\bibitem [{\citenamefont {Casta\~no Yepes}\ and\ \citenamefont
  {Mu\~noz}(2024)}]{Castano-Yepes:2024ctr}%
  \BibitemOpen
  \bibfield  {author} {\bibinfo {author} {\bibfnamefont {J.~D.}\ \bibnamefont
  {Casta\~no Yepes}}\ and\ \bibinfo {author} {\bibfnamefont {E.}~\bibnamefont
  {Mu\~noz}},\ }\href {\doibase 10.1103/PhysRevD.109.056007} {\bibfield
  {journal} {\bibinfo  {journal} {Phys. Rev. D}\ }\textbf {\bibinfo {volume}
  {109}},\ \bibinfo {pages} {056007} (\bibinfo {year} {2024})},\ \Eprint
  {http://arxiv.org/abs/2401.12401} {arXiv:2401.12401 [hep-th]} \BibitemShut
  {NoStop}%
\bibitem [{\citenamefont {Wang}\ \emph {et~al.}(2022)\citenamefont {Wang},
  \citenamefont {Zhao}, \citenamefont {Greiner}, \citenamefont {Xu},\ and\
  \citenamefont {Zhuang}}]{Wang:2021oqq}%
  \BibitemOpen
  \bibfield  {author} {\bibinfo {author} {\bibfnamefont {Z.}~\bibnamefont
  {Wang}}, \bibinfo {author} {\bibfnamefont {J.}~\bibnamefont {Zhao}}, \bibinfo
  {author} {\bibfnamefont {C.}~\bibnamefont {Greiner}}, \bibinfo {author}
  {\bibfnamefont {Z.}~\bibnamefont {Xu}}, \ and\ \bibinfo {author}
  {\bibfnamefont {P.}~\bibnamefont {Zhuang}},\ }\href {\doibase
  10.1103/PhysRevC.105.L041901} {\bibfield  {journal} {\bibinfo  {journal}
  {Phys. Rev. C}\ }\textbf {\bibinfo {volume} {105}},\ \bibinfo {pages}
  {L041901} (\bibinfo {year} {2022})},\ \Eprint
  {http://arxiv.org/abs/2110.14302} {arXiv:2110.14302 [hep-ph]} \BibitemShut
  {NoStop}%
\bibitem [{\citenamefont {Shovkovy}(2022)}]{Shovkovy:2022bnd}%
  \BibitemOpen
  \bibfield  {author} {\bibinfo {author} {\bibfnamefont {I.~A.}\ \bibnamefont
  {Shovkovy}},\ }\href {\doibase 10.3390/particles5040034} {\bibfield
  {journal} {\bibinfo  {journal} {Particles}\ }\textbf {\bibinfo {volume}
  {5}},\ \bibinfo {pages} {442} (\bibinfo {year} {2022})},\ \Eprint
  {http://arxiv.org/abs/2210.00691} {arXiv:2210.00691 [nucl-th]} \BibitemShut
  {NoStop}%
\bibitem [{\citenamefont {Aboona}\ \emph {et~al.}(2023)\citenamefont {Aboona}
  \emph {et~al.}}]{STAR:2022ahj}%
  \BibitemOpen
  \bibfield  {author} {\bibinfo {author} {\bibfnamefont {B.}~\bibnamefont
  {Aboona}} \emph {et~al.} (\bibinfo {collaboration} {STAR}),\ }\href {\doibase
  10.1016/j.physletb.2023.137779} {\bibfield  {journal} {\bibinfo  {journal}
  {Phys. Lett. B}\ }\textbf {\bibinfo {volume} {839}},\ \bibinfo {pages}
  {137779} (\bibinfo {year} {2023})},\ \Eprint
  {http://arxiv.org/abs/2209.03467} {arXiv:2209.03467 [nucl-ex]} \BibitemShut
  {NoStop}%
\bibitem [{\citenamefont {Zhao}\ and\ \citenamefont {Ma}(2022)}]{Zhao:2022grq}%
  \BibitemOpen
  \bibfield  {author} {\bibinfo {author} {\bibfnamefont {X.-L.}\ \bibnamefont
  {Zhao}}\ and\ \bibinfo {author} {\bibfnamefont {G.-L.}\ \bibnamefont {Ma}},\
  }\href {\doibase 10.1103/PhysRevC.106.034909} {\bibfield  {journal} {\bibinfo
   {journal} {Phys. Rev. C}\ }\textbf {\bibinfo {volume} {106}},\ \bibinfo
  {pages} {034909} (\bibinfo {year} {2022})},\ \Eprint
  {http://arxiv.org/abs/2203.15214} {arXiv:2203.15214 [nucl-th]} \BibitemShut
  {NoStop}%
\bibitem [{\citenamefont {Adamczyk}\ \emph {et~al.}(2014)\citenamefont
  {Adamczyk} \emph {et~al.}}]{STAR:2014uiw}%
  \BibitemOpen
  \bibfield  {author} {\bibinfo {author} {\bibfnamefont {L.}~\bibnamefont
  {Adamczyk}} \emph {et~al.} (\bibinfo {collaboration} {STAR}),\ }\href
  {\doibase 10.1103/PhysRevLett.113.052302} {\bibfield  {journal} {\bibinfo
  {journal} {Phys. Rev. Lett.}\ }\textbf {\bibinfo {volume} {113}},\ \bibinfo
  {pages} {052302} (\bibinfo {year} {2014})},\ \Eprint
  {http://arxiv.org/abs/1404.1433} {arXiv:1404.1433 [nucl-ex]} \BibitemShut
  {NoStop}%
\bibitem [{\citenamefont {Kharzeev}\ \emph {et~al.}(2008)\citenamefont
  {Kharzeev}, \citenamefont {McLerran},\ and\ \citenamefont
  {Warringa}}]{Kharzeev:2007jp}%
  \BibitemOpen
  \bibfield  {author} {\bibinfo {author} {\bibfnamefont {D.~E.}\ \bibnamefont
  {Kharzeev}}, \bibinfo {author} {\bibfnamefont {L.~D.}\ \bibnamefont
  {McLerran}}, \ and\ \bibinfo {author} {\bibfnamefont {H.~J.}\ \bibnamefont
  {Warringa}},\ }\href {\doibase 10.1016/j.nuclphysa.2008.02.298} {\bibfield
  {journal} {\bibinfo  {journal} {Nucl. Phys. A}\ }\textbf {\bibinfo {volume}
  {803}},\ \bibinfo {pages} {227} (\bibinfo {year} {2008})},\ \Eprint
  {http://arxiv.org/abs/0711.0950} {arXiv:0711.0950 [hep-ph]} \BibitemShut
  {NoStop}%
\bibitem [{\citenamefont {Adare}\ \emph {et~al.}(2012)\citenamefont {Adare}
  \emph {et~al.}}]{PHENIX:2011oxq}%
  \BibitemOpen
  \bibfield  {author} {\bibinfo {author} {\bibfnamefont {A.}~\bibnamefont
  {Adare}} \emph {et~al.} (\bibinfo {collaboration} {PHENIX}),\ }\href
  {\doibase 10.1103/PhysRevLett.109.122302} {\bibfield  {journal} {\bibinfo
  {journal} {Phys. Rev. Lett.}\ }\textbf {\bibinfo {volume} {109}},\ \bibinfo
  {pages} {122302} (\bibinfo {year} {2012})},\ \Eprint
  {http://arxiv.org/abs/1105.4126} {arXiv:1105.4126 [nucl-ex]} \BibitemShut
  {NoStop}%
\bibitem [{\citenamefont {Acharya}\ \emph
  {et~al.}(2019{\natexlab{a}})\citenamefont {Acharya} \emph
  {et~al.}}]{ALICE:2018mjj}%
  \BibitemOpen
  \bibfield  {author} {\bibinfo {author} {\bibfnamefont {S.}~\bibnamefont
  {Acharya}} \emph {et~al.} (\bibinfo {collaboration} {ALICE}),\ }\href
  {\doibase 10.1103/PhysRevC.99.024912} {\bibfield  {journal} {\bibinfo
  {journal} {Phys. Rev. C}\ }\textbf {\bibinfo {volume} {99}},\ \bibinfo
  {pages} {024912} (\bibinfo {year} {2019}{\natexlab{a}})},\ \Eprint
  {http://arxiv.org/abs/1803.09857} {arXiv:1803.09857 [nucl-ex]} \BibitemShut
  {NoStop}%
\bibitem [{\citenamefont {Adare}\ \emph {et~al.}(2016)\citenamefont {Adare}
  \emph {et~al.}}]{PHENIX:2015igl}%
  \BibitemOpen
  \bibfield  {author} {\bibinfo {author} {\bibfnamefont {A.}~\bibnamefont
  {Adare}} \emph {et~al.} (\bibinfo {collaboration} {PHENIX}),\ }\href
  {\doibase 10.1103/PhysRevC.94.064901} {\bibfield  {journal} {\bibinfo
  {journal} {Phys. Rev. C}\ }\textbf {\bibinfo {volume} {94}},\ \bibinfo
  {pages} {064901} (\bibinfo {year} {2016})},\ \Eprint
  {http://arxiv.org/abs/1509.07758} {arXiv:1509.07758 [nucl-ex]} \BibitemShut
  {NoStop}%
\bibitem [{\citenamefont {Acharya}\ \emph
  {et~al.}(2019{\natexlab{b}})\citenamefont {Acharya} \emph
  {et~al.}}]{ALICE:2018dti}%
  \BibitemOpen
  \bibfield  {author} {\bibinfo {author} {\bibfnamefont {S.}~\bibnamefont
  {Acharya}} \emph {et~al.} (\bibinfo {collaboration} {ALICE}),\ }\href
  {\doibase 10.1016/j.physletb.2018.11.039} {\bibfield  {journal} {\bibinfo
  {journal} {Phys. Lett. B}\ }\textbf {\bibinfo {volume} {789}},\ \bibinfo
  {pages} {308} (\bibinfo {year} {2019}{\natexlab{b}})},\ \Eprint
  {http://arxiv.org/abs/1805.04403} {arXiv:1805.04403 [nucl-ex]} \BibitemShut
  {NoStop}%
\bibitem [{\citenamefont {David}(2020)}]{David:2019wpt}%
  \BibitemOpen
  \bibfield  {author} {\bibinfo {author} {\bibfnamefont {G.}~\bibnamefont
  {David}},\ }\href {\doibase 10.1088/1361-6633/ab6f57} {\bibfield  {journal}
  {\bibinfo  {journal} {Rept. Prog. Phys.}\ }\textbf {\bibinfo {volume} {83}},\
  \bibinfo {pages} {046301} (\bibinfo {year} {2020})},\ \Eprint
  {http://arxiv.org/abs/1907.08893} {arXiv:1907.08893 [nucl-ex]} \BibitemShut
  {NoStop}%
\bibitem [{\citenamefont {Adare}\ \emph {et~al.}(2015)\citenamefont {Adare}
  \emph {et~al.}}]{PHENIX:2014nkk}%
  \BibitemOpen
  \bibfield  {author} {\bibinfo {author} {\bibfnamefont {A.}~\bibnamefont
  {Adare}} \emph {et~al.} (\bibinfo {collaboration} {PHENIX}),\ }\href
  {\doibase 10.1103/PhysRevC.91.064904} {\bibfield  {journal} {\bibinfo
  {journal} {Phys. Rev. C}\ }\textbf {\bibinfo {volume} {91}},\ \bibinfo
  {pages} {064904} (\bibinfo {year} {2015})},\ \Eprint
  {http://arxiv.org/abs/1405.3940} {arXiv:1405.3940 [nucl-ex]} \BibitemShut
  {NoStop}%
\bibitem [{\citenamefont {Adamczyk}\ \emph {et~al.}(2017)\citenamefont
  {Adamczyk} \emph {et~al.}}]{STAR:2016use}%
  \BibitemOpen
  \bibfield  {author} {\bibinfo {author} {\bibfnamefont {L.}~\bibnamefont
  {Adamczyk}} \emph {et~al.} (\bibinfo {collaboration} {STAR}),\ }\href
  {\doibase 10.1016/j.physletb.2017.04.050} {\bibfield  {journal} {\bibinfo
  {journal} {Phys. Lett. B}\ }\textbf {\bibinfo {volume} {770}},\ \bibinfo
  {pages} {451} (\bibinfo {year} {2017})},\ \Eprint
  {http://arxiv.org/abs/1607.01447} {arXiv:1607.01447 [nucl-ex]} \BibitemShut
  {NoStop}%
\bibitem [{\citenamefont {Adam}\ \emph {et~al.}(2016)\citenamefont {Adam} \emph
  {et~al.}}]{ALICE:2015xmh}%
  \BibitemOpen
  \bibfield  {author} {\bibinfo {author} {\bibfnamefont {J.}~\bibnamefont
  {Adam}} \emph {et~al.} (\bibinfo {collaboration} {ALICE}),\ }\href {\doibase
  10.1016/j.physletb.2016.01.020} {\bibfield  {journal} {\bibinfo  {journal}
  {Phys. Lett. B}\ }\textbf {\bibinfo {volume} {754}},\ \bibinfo {pages} {235}
  (\bibinfo {year} {2016})},\ \Eprint {http://arxiv.org/abs/1509.07324}
  {arXiv:1509.07324 [nucl-ex]} \BibitemShut {NoStop}%
\bibitem [{\citenamefont {Adare}\ \emph {et~al.}(2018)\citenamefont {Adare}
  \emph {et~al.}}]{PHENIX:2018che}%
  \BibitemOpen
  \bibfield  {author} {\bibinfo {author} {\bibfnamefont {A.}~\bibnamefont
  {Adare}} \emph {et~al.} (\bibinfo {collaboration} {PHENIX}),\ }\href
  {\doibase 10.1103/PhysRevC.98.054902} {\bibfield  {journal} {\bibinfo
  {journal} {Phys. Rev. C}\ }\textbf {\bibinfo {volume} {98}},\ \bibinfo
  {pages} {054902} (\bibinfo {year} {2018})},\ \Eprint
  {http://arxiv.org/abs/1805.04066} {arXiv:1805.04066 [hep-ex]} \BibitemShut
  {NoStop}%
\bibitem [{\citenamefont {Gale}\ \emph {et~al.}(2022)\citenamefont {Gale},
  \citenamefont {Paquet}, \citenamefont {Schenke},\ and\ \citenamefont
  {Shen}}]{Gale:2021emg}%
  \BibitemOpen
  \bibfield  {author} {\bibinfo {author} {\bibfnamefont {C.}~\bibnamefont
  {Gale}}, \bibinfo {author} {\bibfnamefont {J.-F.}\ \bibnamefont {Paquet}},
  \bibinfo {author} {\bibfnamefont {B.}~\bibnamefont {Schenke}}, \ and\
  \bibinfo {author} {\bibfnamefont {C.}~\bibnamefont {Shen}},\ }\href {\doibase
  10.1103/PhysRevC.105.014909} {\bibfield  {journal} {\bibinfo  {journal}
  {Phys. Rev. C}\ }\textbf {\bibinfo {volume} {105}},\ \bibinfo {pages}
  {014909} (\bibinfo {year} {2022})},\ \Eprint
  {http://arxiv.org/abs/2106.11216} {arXiv:2106.11216 [nucl-th]} \BibitemShut
  {NoStop}%
\bibitem [{\citenamefont {Abgaryan}\ \emph {et~al.}(2022)\citenamefont
  {Abgaryan} \emph {et~al.}}]{MPD:2022qhn}%
  \BibitemOpen
  \bibfield  {author} {\bibinfo {author} {\bibfnamefont {V.}~\bibnamefont
  {Abgaryan}} \emph {et~al.} (\bibinfo {collaboration} {MPD}),\ }\href@noop {}
  {\  (\bibinfo {year} {2022})},\ \Eprint {http://arxiv.org/abs/2202.08970}
  {arXiv:2202.08970 [physics.ins-det]} \BibitemShut {NoStop}%
\bibitem [{\citenamefont {Tuchin}(2015)}]{Tuchin:2014pka}%
  \BibitemOpen
  \bibfield  {author} {\bibinfo {author} {\bibfnamefont {K.}~\bibnamefont
  {Tuchin}},\ }\href {\doibase 10.1103/PhysRevC.91.014902} {\bibfield
  {journal} {\bibinfo  {journal} {Phys. Rev. C}\ }\textbf {\bibinfo {volume}
  {91}},\ \bibinfo {pages} {014902} (\bibinfo {year} {2015})},\ \Eprint
  {http://arxiv.org/abs/1406.5097} {arXiv:1406.5097 [nucl-th]} \BibitemShut
  {NoStop}%
\bibitem [{\citenamefont {Zakharov}(2016)}]{Zakharov:2016mmc}%
  \BibitemOpen
  \bibfield  {author} {\bibinfo {author} {\bibfnamefont {B.~G.}\ \bibnamefont
  {Zakharov}},\ }\href {\doibase 10.1140/epjc/s10052-016-4451-8} {\bibfield
  {journal} {\bibinfo  {journal} {Eur. Phys. J. C}\ }\textbf {\bibinfo {volume}
  {76}},\ \bibinfo {pages} {609} (\bibinfo {year} {2016})},\ \Eprint
  {http://arxiv.org/abs/1609.04324} {arXiv:1609.04324 [nucl-th]} \BibitemShut
  {NoStop}%
\bibitem [{\citenamefont {Wang}\ and\ \citenamefont
  {Shovkovy}(2022)}]{Wang:2022jxx}%
  \BibitemOpen
  \bibfield  {author} {\bibinfo {author} {\bibfnamefont {X.}~\bibnamefont
  {Wang}}\ and\ \bibinfo {author} {\bibfnamefont {I.~A.}\ \bibnamefont
  {Shovkovy}},\ }\href {\doibase 10.1103/PhysRevD.106.036014} {\bibfield
  {journal} {\bibinfo  {journal} {Phys. Rev. D}\ }\textbf {\bibinfo {volume}
  {106}},\ \bibinfo {pages} {036014} (\bibinfo {year} {2022})},\ \Eprint
  {http://arxiv.org/abs/2205.00276} {arXiv:2205.00276 [nucl-th]} \BibitemShut
  {NoStop}%
\bibitem [{\citenamefont {Wang}\ \emph {et~al.}(2020)\citenamefont {Wang},
  \citenamefont {Shovkovy}, \citenamefont {Yu},\ and\ \citenamefont
  {Huang}}]{Wang:2020dsr}%
  \BibitemOpen
  \bibfield  {author} {\bibinfo {author} {\bibfnamefont {X.}~\bibnamefont
  {Wang}}, \bibinfo {author} {\bibfnamefont {I.~A.}\ \bibnamefont {Shovkovy}},
  \bibinfo {author} {\bibfnamefont {L.}~\bibnamefont {Yu}}, \ and\ \bibinfo
  {author} {\bibfnamefont {M.}~\bibnamefont {Huang}},\ }\href {\doibase
  10.1103/PhysRevD.102.076010} {\bibfield  {journal} {\bibinfo  {journal}
  {Phys. Rev. D}\ }\textbf {\bibinfo {volume} {102}},\ \bibinfo {pages}
  {076010} (\bibinfo {year} {2020})},\ \Eprint
  {http://arxiv.org/abs/2006.16254} {arXiv:2006.16254 [hep-ph]} \BibitemShut
  {NoStop}%
\bibitem [{\citenamefont {Buzzegoli}\ \emph {et~al.}(2023)\citenamefont
  {Buzzegoli}, \citenamefont {Kroth}, \citenamefont {Tuchin},\ and\
  \citenamefont {Vijayakumar}}]{Buzzegoli:2023vne}%
  \BibitemOpen
  \bibfield  {author} {\bibinfo {author} {\bibfnamefont {M.}~\bibnamefont
  {Buzzegoli}}, \bibinfo {author} {\bibfnamefont {J.~D.}\ \bibnamefont
  {Kroth}}, \bibinfo {author} {\bibfnamefont {K.}~\bibnamefont {Tuchin}}, \
  and\ \bibinfo {author} {\bibfnamefont {N.}~\bibnamefont {Vijayakumar}},\
  }\href {\doibase 10.1103/PhysRevD.108.096014} {\bibfield  {journal} {\bibinfo
   {journal} {Phys. Rev. D}\ }\textbf {\bibinfo {volume} {108}},\ \bibinfo
  {pages} {096014} (\bibinfo {year} {2023})},\ \Eprint
  {http://arxiv.org/abs/2306.03863} {arXiv:2306.03863 [hep-ph]} \BibitemShut
  {NoStop}%
\bibitem [{\citenamefont {Wang}\ and\ \citenamefont
  {Shovkovy}(2021)}]{Wang:2021eud}%
  \BibitemOpen
  \bibfield  {author} {\bibinfo {author} {\bibfnamefont {X.}~\bibnamefont
  {Wang}}\ and\ \bibinfo {author} {\bibfnamefont {I.}~\bibnamefont
  {Shovkovy}},\ }\href {\doibase 10.1140/epjc/s10052-021-09650-3} {\bibfield
  {journal} {\bibinfo  {journal} {Eur. Phys. J. C}\ }\textbf {\bibinfo {volume}
  {81}},\ \bibinfo {pages} {901} (\bibinfo {year} {2021})},\ \Eprint
  {http://arxiv.org/abs/2106.09029} {arXiv:2106.09029 [nucl-th]} \BibitemShut
  {NoStop}%
\bibitem [{\citenamefont {Sadooghi}\ and\ \citenamefont
  {Taghinavaz}(2017)}]{Sadooghi:2016jyf}%
  \BibitemOpen
  \bibfield  {author} {\bibinfo {author} {\bibfnamefont {N.}~\bibnamefont
  {Sadooghi}}\ and\ \bibinfo {author} {\bibfnamefont {F.}~\bibnamefont
  {Taghinavaz}},\ }\href {\doibase 10.1016/j.aop.2016.11.008} {\bibfield
  {journal} {\bibinfo  {journal} {Annals Phys.}\ }\textbf {\bibinfo {volume}
  {376}},\ \bibinfo {pages} {218} (\bibinfo {year} {2017})},\ \Eprint
  {http://arxiv.org/abs/1601.04887} {arXiv:1601.04887 [hep-ph]} \BibitemShut
  {NoStop}%
\bibitem [{\citenamefont {Mondal}\ \emph
  {et~al.}(2023{\natexlab{a}})\citenamefont {Mondal}, \citenamefont
  {Chaudhuri}, \citenamefont {Ghosh}, \citenamefont {Sarkar},\ and\
  \citenamefont {Roy}}]{Mondal:2023vzx}%
  \BibitemOpen
  \bibfield  {author} {\bibinfo {author} {\bibfnamefont {R.}~\bibnamefont
  {Mondal}}, \bibinfo {author} {\bibfnamefont {N.}~\bibnamefont {Chaudhuri}},
  \bibinfo {author} {\bibfnamefont {S.}~\bibnamefont {Ghosh}}, \bibinfo
  {author} {\bibfnamefont {S.}~\bibnamefont {Sarkar}}, \ and\ \bibinfo {author}
  {\bibfnamefont {P.}~\bibnamefont {Roy}},\ }\href {\doibase
  10.1103/PhysRevD.107.036017} {\bibfield  {journal} {\bibinfo  {journal}
  {Phys. Rev. D}\ }\textbf {\bibinfo {volume} {107}},\ \bibinfo {pages}
  {036017} (\bibinfo {year} {2023}{\natexlab{a}})},\ \Eprint
  {http://arxiv.org/abs/2301.09475} {arXiv:2301.09475 [hep-ph]} \BibitemShut
  {NoStop}%
\bibitem [{\citenamefont {Mondal}\ \emph
  {et~al.}(2023{\natexlab{b}})\citenamefont {Mondal}, \citenamefont
  {Chaudhuri}, \citenamefont {Ghosh}, \citenamefont {Sarkar},\ and\
  \citenamefont {Roy}}]{Mondal:2023ypq}%
  \BibitemOpen
  \bibfield  {author} {\bibinfo {author} {\bibfnamefont {R.}~\bibnamefont
  {Mondal}}, \bibinfo {author} {\bibfnamefont {N.}~\bibnamefont {Chaudhuri}},
  \bibinfo {author} {\bibfnamefont {S.}~\bibnamefont {Ghosh}}, \bibinfo
  {author} {\bibfnamefont {S.}~\bibnamefont {Sarkar}}, \ and\ \bibinfo {author}
  {\bibfnamefont {P.}~\bibnamefont {Roy}},\ }\href {\doibase
  10.1140/epja/s10050-023-01201-6} {\bibfield  {journal} {\bibinfo  {journal}
  {Eur. Phys. J. A}\ }\textbf {\bibinfo {volume} {59}},\ \bibinfo {pages} {287}
  (\bibinfo {year} {2023}{\natexlab{b}})},\ \Eprint
  {http://arxiv.org/abs/2311.17632} {arXiv:2311.17632 [hep-ph]} \BibitemShut
  {NoStop}%
\bibitem [{\citenamefont {Basar}\ \emph {et~al.}(2012)\citenamefont {Basar},
  \citenamefont {Kharzeev},\ and\ \citenamefont {Skokov}}]{Basar:2012bp}%
  \BibitemOpen
  \bibfield  {author} {\bibinfo {author} {\bibfnamefont {G.}~\bibnamefont
  {Basar}}, \bibinfo {author} {\bibfnamefont {D.}~\bibnamefont {Kharzeev}}, \
  and\ \bibinfo {author} {\bibfnamefont {V.}~\bibnamefont {Skokov}},\ }\href
  {\doibase 10.1103/PhysRevLett.109.202303} {\bibfield  {journal} {\bibinfo
  {journal} {Phys. Rev. Lett.}\ }\textbf {\bibinfo {volume} {109}},\ \bibinfo
  {pages} {202303} (\bibinfo {year} {2012})},\ \Eprint
  {http://arxiv.org/abs/1206.1334} {arXiv:1206.1334 [hep-ph]} \BibitemShut
  {NoStop}%
\bibitem [{\citenamefont {Basar}\ \emph {et~al.}(2014)\citenamefont {Basar},
  \citenamefont {Kharzeev},\ and\ \citenamefont {Shuryak}}]{Basar:2014swa}%
  \BibitemOpen
  \bibfield  {author} {\bibinfo {author} {\bibfnamefont {G.}~\bibnamefont
  {Basar}}, \bibinfo {author} {\bibfnamefont {D.~E.}\ \bibnamefont {Kharzeev}},
  \ and\ \bibinfo {author} {\bibfnamefont {E.~V.}\ \bibnamefont {Shuryak}},\
  }\href {\doibase 10.1103/PhysRevC.90.014905} {\bibfield  {journal} {\bibinfo
  {journal} {Phys. Rev. C}\ }\textbf {\bibinfo {volume} {90}},\ \bibinfo
  {pages} {014905} (\bibinfo {year} {2014})},\ \Eprint
  {http://arxiv.org/abs/1402.2286} {arXiv:1402.2286 [hep-ph]} \BibitemShut
  {NoStop}%
\bibitem [{\citenamefont {Lee}(2020)}]{Lee:2020tay}%
  \BibitemOpen
  \bibfield  {author} {\bibinfo {author} {\bibfnamefont {C.-Y.}\ \bibnamefont
  {Lee}},\ }\href {\doibase 10.1016/j.physletb.2020.135794} {\bibfield
  {journal} {\bibinfo  {journal} {Phys. Lett. B}\ }\textbf {\bibinfo {volume}
  {810}},\ \bibinfo {pages} {135794} (\bibinfo {year} {2020})},\ \Eprint
  {http://arxiv.org/abs/2005.11657} {arXiv:2005.11657 [hep-ph]} \BibitemShut
  {NoStop}%
\bibitem [{\citenamefont {Sun}\ and\ \citenamefont
  {Yan}(2023{\natexlab{b}})}]{Sun:2023pil}%
  \BibitemOpen
  \bibfield  {author} {\bibinfo {author} {\bibfnamefont {J.-A.}\ \bibnamefont
  {Sun}}\ and\ \bibinfo {author} {\bibfnamefont {L.}~\bibnamefont {Yan}},\
  }\href@noop {} {\  (\bibinfo {year} {2023}{\natexlab{b}})},\ \Eprint
  {http://arxiv.org/abs/2302.07696} {arXiv:2302.07696 [nucl-th]} \BibitemShut
  {NoStop}%
\bibitem [{\citenamefont {\'Avila}\ \emph {et~al.}(2023)\citenamefont
  {\'Avila}, \citenamefont {Nettel},\ and\ \citenamefont
  {Pati\~no}}]{Avila:2022cpa}%
  \BibitemOpen
  \bibfield  {author} {\bibinfo {author} {\bibfnamefont {D.}~\bibnamefont
  {\'Avila}}, \bibinfo {author} {\bibfnamefont {F.}~\bibnamefont {Nettel}}, \
  and\ \bibinfo {author} {\bibfnamefont {L.}~\bibnamefont {Pati\~no}},\ }\href
  {\doibase 10.1103/PhysRevD.107.066010} {\bibfield  {journal} {\bibinfo
  {journal} {Phys. Rev. D}\ }\textbf {\bibinfo {volume} {107}},\ \bibinfo
  {pages} {066010} (\bibinfo {year} {2023})},\ \Eprint
  {http://arxiv.org/abs/2204.00024} {arXiv:2204.00024 [hep-th]} \BibitemShut
  {NoStop}%
\bibitem [{\citenamefont {Arciniega}\ \emph {et~al.}(2014)\citenamefont
  {Arciniega}, \citenamefont {Nettel}, \citenamefont {Ortega},\ and\
  \citenamefont {Pati\~no}}]{Arciniega:2013dqa}%
  \BibitemOpen
  \bibfield  {author} {\bibinfo {author} {\bibfnamefont {G.}~\bibnamefont
  {Arciniega}}, \bibinfo {author} {\bibfnamefont {F.}~\bibnamefont {Nettel}},
  \bibinfo {author} {\bibfnamefont {P.}~\bibnamefont {Ortega}}, \ and\ \bibinfo
  {author} {\bibfnamefont {L.}~\bibnamefont {Pati\~no}},\ }\href {\doibase
  10.1007/JHEP04(2014)192} {\bibfield  {journal} {\bibinfo  {journal} {JHEP}\
  }\textbf {\bibinfo {volume} {04}},\ \bibinfo {pages} {192} (\bibinfo {year}
  {2014})},\ \Eprint {http://arxiv.org/abs/1307.1153} {arXiv:1307.1153
  [hep-th]} \BibitemShut {NoStop}%
\bibitem [{\citenamefont {Mamo}(2013)}]{Mamo:2012kqw}%
  \BibitemOpen
  \bibfield  {author} {\bibinfo {author} {\bibfnamefont {K.~A.}\ \bibnamefont
  {Mamo}},\ }\href {\doibase 10.1007/JHEP08(2013)083} {\bibfield  {journal}
  {\bibinfo  {journal} {JHEP}\ }\textbf {\bibinfo {volume} {08}},\ \bibinfo
  {pages} {083} (\bibinfo {year} {2013})},\ \Eprint
  {http://arxiv.org/abs/1210.7428} {arXiv:1210.7428 [hep-th]} \BibitemShut
  {NoStop}%
\bibitem [{\citenamefont {Wu}\ and\ \citenamefont {Yang}(2013)}]{Wu:2013qja}%
  \BibitemOpen
  \bibfield  {author} {\bibinfo {author} {\bibfnamefont {S.-Y.}\ \bibnamefont
  {Wu}}\ and\ \bibinfo {author} {\bibfnamefont {D.-L.}\ \bibnamefont {Yang}},\
  }\href {\doibase 10.1007/JHEP08(2013)032} {\bibfield  {journal} {\bibinfo
  {journal} {JHEP}\ }\textbf {\bibinfo {volume} {08}},\ \bibinfo {pages} {032}
  (\bibinfo {year} {2013})},\ \Eprint {http://arxiv.org/abs/1305.5509}
  {arXiv:1305.5509 [hep-th]} \BibitemShut {NoStop}%
\bibitem [{\citenamefont {Nedelko}\ and\ \citenamefont
  {Nikolskii}(2023)}]{Nedelko:2022kjy}%
  \BibitemOpen
  \bibfield  {author} {\bibinfo {author} {\bibfnamefont {S.}~\bibnamefont
  {Nedelko}}\ and\ \bibinfo {author} {\bibfnamefont {A.}~\bibnamefont
  {Nikolskii}},\ }\href {\doibase 10.1140/epja/s10050-023-00986-w} {\bibfield
  {journal} {\bibinfo  {journal} {Eur. Phys. J. A}\ }\textbf {\bibinfo {volume}
  {59}},\ \bibinfo {pages} {70} (\bibinfo {year} {2023})},\ \Eprint
  {http://arxiv.org/abs/2208.00842} {arXiv:2208.00842 [hep-ph]} \BibitemShut
  {NoStop}%
\bibitem [{\citenamefont {Ayala}\ \emph {et~al.}(2017)\citenamefont {Ayala},
  \citenamefont {Castano-Yepes}, \citenamefont {Dominguez}, \citenamefont
  {Hernandez}, \citenamefont {Hernandez-Ortiz},\ and\ \citenamefont
  {Tejeda-Yeomans}}]{Ayala:2017vex}%
  \BibitemOpen
  \bibfield  {author} {\bibinfo {author} {\bibfnamefont {A.}~\bibnamefont
  {Ayala}}, \bibinfo {author} {\bibfnamefont {J.~D.}\ \bibnamefont
  {Castano-Yepes}}, \bibinfo {author} {\bibfnamefont {C.~A.}\ \bibnamefont
  {Dominguez}}, \bibinfo {author} {\bibfnamefont {L.~A.}\ \bibnamefont
  {Hernandez}}, \bibinfo {author} {\bibfnamefont {S.}~\bibnamefont
  {Hernandez-Ortiz}}, \ and\ \bibinfo {author} {\bibfnamefont {M.~E.}\
  \bibnamefont {Tejeda-Yeomans}},\ }\href {\doibase 10.1103/PhysRevD.96.014023}
  {\bibfield  {journal} {\bibinfo  {journal} {Phys. Rev. D}\ }\textbf {\bibinfo
  {volume} {96}},\ \bibinfo {pages} {014023} (\bibinfo {year} {2017})},\
  \bibinfo {note} {[Erratum: Phys.Rev.D 96, 119901 (2017)]},\ \Eprint
  {http://arxiv.org/abs/1704.02433} {arXiv:1704.02433 [hep-ph]} \BibitemShut
  {NoStop}%
\bibitem [{\citenamefont {Ayala}\ \emph
  {et~al.}(2020{\natexlab{a}})\citenamefont {Ayala}, \citenamefont {Casta\~no
  Yepes}, \citenamefont {Dominguez~Jimenez}, \citenamefont {Salinas
  San~Mart\'\i{}n},\ and\ \citenamefont {Tejeda-Yeomans}}]{Ayala:2019jey}%
  \BibitemOpen
  \bibfield  {author} {\bibinfo {author} {\bibfnamefont {A.}~\bibnamefont
  {Ayala}}, \bibinfo {author} {\bibfnamefont {J.~D.}\ \bibnamefont {Casta\~no
  Yepes}}, \bibinfo {author} {\bibfnamefont {I.}~\bibnamefont
  {Dominguez~Jimenez}}, \bibinfo {author} {\bibfnamefont {J.}~\bibnamefont
  {Salinas San~Mart\'\i{}n}}, \ and\ \bibinfo {author} {\bibfnamefont {M.~E.}\
  \bibnamefont {Tejeda-Yeomans}},\ }\href {\doibase
  10.1140/epja/s10050-020-00060-9} {\bibfield  {journal} {\bibinfo  {journal}
  {Eur. Phys. J. A}\ }\textbf {\bibinfo {volume} {56}},\ \bibinfo {pages} {53}
  (\bibinfo {year} {2020}{\natexlab{a}})},\ \Eprint
  {http://arxiv.org/abs/1904.02938} {arXiv:1904.02938 [hep-ph]} \BibitemShut
  {NoStop}%
\bibitem [{\citenamefont {Ayala}\ \emph {et~al.}(2022)\citenamefont {Ayala},
  \citenamefont {Casta\~no Yepes}, \citenamefont {Hern\'andez}, \citenamefont
  {Mizher}, \citenamefont {Tejeda-Yeomans},\ and\ \citenamefont
  {Zamora}}]{Ayala:2022zhu}%
  \BibitemOpen
  \bibfield  {author} {\bibinfo {author} {\bibfnamefont {A.}~\bibnamefont
  {Ayala}}, \bibinfo {author} {\bibfnamefont {J.~D.}\ \bibnamefont {Casta\~no
  Yepes}}, \bibinfo {author} {\bibfnamefont {L.~A.}\ \bibnamefont
  {Hern\'andez}}, \bibinfo {author} {\bibfnamefont {A.~J.}\ \bibnamefont
  {Mizher}}, \bibinfo {author} {\bibfnamefont {M.~E.}\ \bibnamefont
  {Tejeda-Yeomans}}, \ and\ \bibinfo {author} {\bibfnamefont {R.}~\bibnamefont
  {Zamora}},\ }\href {\doibase 10.1103/PhysRevC.106.064905} {\bibfield
  {journal} {\bibinfo  {journal} {Phys. Rev. C}\ }\textbf {\bibinfo {volume}
  {106}},\ \bibinfo {pages} {064905} (\bibinfo {year} {2022})},\ \Eprint
  {http://arxiv.org/abs/2209.09364} {arXiv:2209.09364 [hep-ph]} \BibitemShut
  {NoStop}%
\bibitem [{\citenamefont {Baier}\ \emph {et~al.}(2001)\citenamefont {Baier},
  \citenamefont {Mueller}, \citenamefont {Schiff},\ and\ \citenamefont
  {Son}}]{Baier:2000sb}%
  \BibitemOpen
  \bibfield  {author} {\bibinfo {author} {\bibfnamefont {R.}~\bibnamefont
  {Baier}}, \bibinfo {author} {\bibfnamefont {A.~H.}\ \bibnamefont {Mueller}},
  \bibinfo {author} {\bibfnamefont {D.}~\bibnamefont {Schiff}}, \ and\ \bibinfo
  {author} {\bibfnamefont {D.~T.}\ \bibnamefont {Son}},\ }\href {\doibase
  10.1016/S0370-2693(01)00191-5} {\bibfield  {journal} {\bibinfo  {journal}
  {Phys. Lett. B}\ }\textbf {\bibinfo {volume} {502}},\ \bibinfo {pages} {51}
  (\bibinfo {year} {2001})},\ \Eprint {http://arxiv.org/abs/hep-ph/0009237}
  {arXiv:hep-ph/0009237} \BibitemShut {NoStop}%
\bibitem [{\citenamefont {Garcia-Montero}(2022)}]{Garcia-Montero:2019vju}%
  \BibitemOpen
  \bibfield  {author} {\bibinfo {author} {\bibfnamefont {O.}~\bibnamefont
  {Garcia-Montero}},\ }\href {\doibase 10.1016/j.aop.2022.168984} {\bibfield
  {journal} {\bibinfo  {journal} {Annals Phys.}\ }\textbf {\bibinfo {volume}
  {443}},\ \bibinfo {pages} {168984} (\bibinfo {year} {2022})},\ \Eprint
  {http://arxiv.org/abs/1909.12294} {arXiv:1909.12294 [hep-ph]} \BibitemShut
  {NoStop}%
\bibitem [{\citenamefont {Monnai}(2020)}]{Monnai:2019vup}%
  \BibitemOpen
  \bibfield  {author} {\bibinfo {author} {\bibfnamefont {A.}~\bibnamefont
  {Monnai}},\ }\href {\doibase 10.1088/1361-6471/ab8d8c} {\bibfield  {journal}
  {\bibinfo  {journal} {J. Phys. G}\ }\textbf {\bibinfo {volume} {47}},\
  \bibinfo {pages} {075105} (\bibinfo {year} {2020})},\ \Eprint
  {http://arxiv.org/abs/1907.09266} {arXiv:1907.09266 [nucl-th]} \BibitemShut
  {NoStop}%
\bibitem [{\citenamefont {McLerran}\ and\ \citenamefont
  {Schenke}(2014)}]{McLerran:2014hza}%
  \BibitemOpen
  \bibfield  {author} {\bibinfo {author} {\bibfnamefont {L.}~\bibnamefont
  {McLerran}}\ and\ \bibinfo {author} {\bibfnamefont {B.}~\bibnamefont
  {Schenke}},\ }\href {\doibase 10.1016/j.nuclphysa.2014.06.004} {\bibfield
  {journal} {\bibinfo  {journal} {Nucl. Phys. A}\ }\textbf {\bibinfo {volume}
  {929}},\ \bibinfo {pages} {71} (\bibinfo {year} {2014})},\ \Eprint
  {http://arxiv.org/abs/1403.7462} {arXiv:1403.7462 [hep-ph]} \BibitemShut
  {NoStop}%
\bibitem [{\citenamefont {Kurkela}\ \emph
  {et~al.}(2019{\natexlab{a}})\citenamefont {Kurkela}, \citenamefont
  {Mazeliauskas}, \citenamefont {Paquet}, \citenamefont {Schlichting},\ and\
  \citenamefont {Teaney}}]{Kurkela:2018vqr}%
  \BibitemOpen
  \bibfield  {author} {\bibinfo {author} {\bibfnamefont {A.}~\bibnamefont
  {Kurkela}}, \bibinfo {author} {\bibfnamefont {A.}~\bibnamefont
  {Mazeliauskas}}, \bibinfo {author} {\bibfnamefont {J.-F.}\ \bibnamefont
  {Paquet}}, \bibinfo {author} {\bibfnamefont {S.}~\bibnamefont {Schlichting}},
  \ and\ \bibinfo {author} {\bibfnamefont {D.}~\bibnamefont {Teaney}},\ }\href
  {\doibase 10.1103/PhysRevC.99.034910} {\bibfield  {journal} {\bibinfo
  {journal} {Phys. Rev. C}\ }\textbf {\bibinfo {volume} {99}},\ \bibinfo
  {pages} {034910} (\bibinfo {year} {2019}{\natexlab{a}})},\ \Eprint
  {http://arxiv.org/abs/1805.00961} {arXiv:1805.00961 [hep-ph]} \BibitemShut
  {NoStop}%
\bibitem [{\citenamefont {Kurkela}\ \emph
  {et~al.}(2019{\natexlab{b}})\citenamefont {Kurkela}, \citenamefont
  {Mazeliauskas}, \citenamefont {Paquet}, \citenamefont {Schlichting},\ and\
  \citenamefont {Teaney}}]{Kurkela:2018wud}%
  \BibitemOpen
  \bibfield  {author} {\bibinfo {author} {\bibfnamefont {A.}~\bibnamefont
  {Kurkela}}, \bibinfo {author} {\bibfnamefont {A.}~\bibnamefont
  {Mazeliauskas}}, \bibinfo {author} {\bibfnamefont {J.-F.}\ \bibnamefont
  {Paquet}}, \bibinfo {author} {\bibfnamefont {S.}~\bibnamefont {Schlichting}},
  \ and\ \bibinfo {author} {\bibfnamefont {D.}~\bibnamefont {Teaney}},\ }\href
  {\doibase 10.1103/PhysRevLett.122.122302} {\bibfield  {journal} {\bibinfo
  {journal} {Phys. Rev. Lett.}\ }\textbf {\bibinfo {volume} {122}},\ \bibinfo
  {pages} {122302} (\bibinfo {year} {2019}{\natexlab{b}})},\ \Eprint
  {http://arxiv.org/abs/1805.01604} {arXiv:1805.01604 [hep-ph]} \BibitemShut
  {NoStop}%
\bibitem [{\citenamefont {Kasmaei}\ and\ \citenamefont
  {Strickland}(2020)}]{Kasmaei:2019ofu}%
  \BibitemOpen
  \bibfield  {author} {\bibinfo {author} {\bibfnamefont {B.~S.}\ \bibnamefont
  {Kasmaei}}\ and\ \bibinfo {author} {\bibfnamefont {M.}~\bibnamefont
  {Strickland}},\ }\href {\doibase 10.1103/PhysRevD.102.014037} {\bibfield
  {journal} {\bibinfo  {journal} {Phys. Rev. D}\ }\textbf {\bibinfo {volume}
  {102}},\ \bibinfo {pages} {014037} (\bibinfo {year} {2020})},\ \Eprint
  {http://arxiv.org/abs/1911.03370} {arXiv:1911.03370 [hep-ph]} \BibitemShut
  {NoStop}%
\bibitem [{\citenamefont {Churchill}\ \emph {et~al.}(2021)\citenamefont
  {Churchill}, \citenamefont {Yan}, \citenamefont {Jeon},\ and\ \citenamefont
  {Gale}}]{Churchill:2020uvk}%
  \BibitemOpen
  \bibfield  {author} {\bibinfo {author} {\bibfnamefont {J.}~\bibnamefont
  {Churchill}}, \bibinfo {author} {\bibfnamefont {L.}~\bibnamefont {Yan}},
  \bibinfo {author} {\bibfnamefont {S.}~\bibnamefont {Jeon}}, \ and\ \bibinfo
  {author} {\bibfnamefont {C.}~\bibnamefont {Gale}},\ }\href {\doibase
  10.1103/PhysRevC.103.024904} {\bibfield  {journal} {\bibinfo  {journal}
  {Phys. Rev. C}\ }\textbf {\bibinfo {volume} {103}},\ \bibinfo {pages}
  {024904} (\bibinfo {year} {2021})},\ \Eprint
  {http://arxiv.org/abs/2008.02902} {arXiv:2008.02902 [hep-ph]} \BibitemShut
  {NoStop}%
\bibitem [{\citenamefont {Garcia-Montero}\ \emph {et~al.}(2023)\citenamefont
  {Garcia-Montero}, \citenamefont {Mazeliauskas}, \citenamefont {Plaschke},\
  and\ \citenamefont {Schlichting}}]{Garcia-Montero:2023lrd}%
  \BibitemOpen
  \bibfield  {author} {\bibinfo {author} {\bibfnamefont {O.}~\bibnamefont
  {Garcia-Montero}}, \bibinfo {author} {\bibfnamefont {A.}~\bibnamefont
  {Mazeliauskas}}, \bibinfo {author} {\bibfnamefont {P.}~\bibnamefont
  {Plaschke}}, \ and\ \bibinfo {author} {\bibfnamefont {S.}~\bibnamefont
  {Schlichting}},\ }\href@noop {} {\  (\bibinfo {year} {2023})},\ \Eprint
  {http://arxiv.org/abs/2308.09747} {arXiv:2308.09747 [hep-ph]} \BibitemShut
  {NoStop}%
\bibitem [{\citenamefont {Jaber-Urquiza}\ and\ \citenamefont
  {Sanchez}(2023)}]{Jaber-Urquiza:2023swn}%
  \BibitemOpen
  \bibfield  {author} {\bibinfo {author} {\bibfnamefont {J.}~\bibnamefont
  {Jaber-Urquiza}}\ and\ \bibinfo {author} {\bibfnamefont {A.}~\bibnamefont
  {Sanchez}},\ }\href {\doibase 10.1103/PhysRevD.107.116024} {\bibfield
  {journal} {\bibinfo  {journal} {Phys. Rev. D}\ }\textbf {\bibinfo {volume}
  {107}},\ \bibinfo {pages} {116024} (\bibinfo {year} {2023})},\ \Eprint
  {http://arxiv.org/abs/2305.17199} {arXiv:2305.17199 [hep-ph]} \BibitemShut
  {NoStop}%
\bibitem [{\citenamefont {{Panyan}}\ and\ \citenamefont
  {{Ritus}}(1972)}]{Papanyan:1971cvCORR}%
  \BibitemOpen
  \bibfield  {author} {\bibinfo {author} {\bibfnamefont {V.~O.}\ \bibnamefont
  {{Panyan}}}\ and\ \bibinfo {author} {\bibfnamefont {V.~I.}\ \bibnamefont
  {{Ritus}}},\ }\href@noop {} {\bibfield  {journal} {\bibinfo  {journal}
  {Soviet Journal of Experimental and Theoretical Physics}\ }\textbf {\bibinfo
  {volume} {34}},\ \bibinfo {pages} {1195} (\bibinfo {year}
  {1972})}\BibitemShut {NoStop}%
\bibitem [{\citenamefont {{Papanyan}}\ and\ \citenamefont
  {{Ritus}}(1974)}]{Papanyan:1973xaCORR}%
  \BibitemOpen
  \bibfield  {author} {\bibinfo {author} {\bibfnamefont {V.~O.}\ \bibnamefont
  {{Papanyan}}}\ and\ \bibinfo {author} {\bibfnamefont {V.~I.}\ \bibnamefont
  {{Ritus}}},\ }\href@noop {} {\bibfield  {journal} {\bibinfo  {journal}
  {Soviet Journal of Experimental and Theoretical Physics}\ }\textbf {\bibinfo
  {volume} {38}},\ \bibinfo {pages} {879} (\bibinfo {year} {1974})}\BibitemShut
  {NoStop}%
\bibitem [{\citenamefont {Hattori}\ and\ \citenamefont
  {Satow}(2018)}]{Hattori:2017xoo}%
  \BibitemOpen
  \bibfield  {author} {\bibinfo {author} {\bibfnamefont {K.}~\bibnamefont
  {Hattori}}\ and\ \bibinfo {author} {\bibfnamefont {D.}~\bibnamefont
  {Satow}},\ }\href {\doibase 10.1103/PhysRevD.97.014023} {\bibfield  {journal}
  {\bibinfo  {journal} {Phys. Rev. D}\ }\textbf {\bibinfo {volume} {97}},\
  \bibinfo {pages} {014023} (\bibinfo {year} {2018})},\ \Eprint
  {http://arxiv.org/abs/1704.03191} {arXiv:1704.03191 [hep-ph]} \BibitemShut
  {NoStop}%
\bibitem [{\citenamefont {Papanyan}\ and\ \citenamefont
  {Ritus}(1971)}]{Papanyan:1971cv}%
  \BibitemOpen
  \bibfield  {author} {\bibinfo {author} {\bibfnamefont {V.~O.}\ \bibnamefont
  {Papanyan}}\ and\ \bibinfo {author} {\bibfnamefont {V.~I.}\ \bibnamefont
  {Ritus}},\ }\href@noop {} {\bibfield  {journal} {\bibinfo  {journal} {Zh.
  Eksp. Teor. Fiz.}\ }\textbf {\bibinfo {volume} {61}},\ \bibinfo {pages}
  {2231} (\bibinfo {year} {1971})}\BibitemShut {NoStop}%
\bibitem [{\citenamefont {Ayala}\ \emph
  {et~al.}(2020{\natexlab{b}})\citenamefont {Ayala}, \citenamefont
  {Hern\'andez}, \citenamefont {Hern\'andez}, \citenamefont {Farias},\ and\
  \citenamefont {Zamora}}]{Ayala:2020muk}%
  \BibitemOpen
  \bibfield  {author} {\bibinfo {author} {\bibfnamefont {A.}~\bibnamefont
  {Ayala}}, \bibinfo {author} {\bibfnamefont {J.~L.}\ \bibnamefont
  {Hern\'andez}}, \bibinfo {author} {\bibfnamefont {L.~A.}\ \bibnamefont
  {Hern\'andez}}, \bibinfo {author} {\bibfnamefont {R.~L.~S.}\ \bibnamefont
  {Farias}}, \ and\ \bibinfo {author} {\bibfnamefont {R.}~\bibnamefont
  {Zamora}},\ }\href {\doibase 10.1103/PhysRevD.102.114038} {\bibfield
  {journal} {\bibinfo  {journal} {Phys. Rev. D}\ }\textbf {\bibinfo {volume}
  {102}},\ \bibinfo {pages} {114038} (\bibinfo {year} {2020}{\natexlab{b}})},\
  \Eprint {http://arxiv.org/abs/2009.13740} {arXiv:2009.13740 [hep-ph]}
  \BibitemShut {NoStop}%
\bibitem [{\citenamefont {Piccinelli}\ and\ \citenamefont
  {Sanchez}(2017)}]{Piccinelli:2017yvl}%
  \BibitemOpen
  \bibfield  {author} {\bibinfo {author} {\bibfnamefont {G.}~\bibnamefont
  {Piccinelli}}\ and\ \bibinfo {author} {\bibfnamefont {A.}~\bibnamefont
  {Sanchez}},\ }\href {\doibase 10.1103/PhysRevD.96.076014} {\bibfield
  {journal} {\bibinfo  {journal} {Phys. Rev. D}\ }\textbf {\bibinfo {volume}
  {96}},\ \bibinfo {pages} {076014} (\bibinfo {year} {2017})},\ \Eprint
  {http://arxiv.org/abs/1707.08257} {arXiv:1707.08257 [hep-ph]} \BibitemShut
  {NoStop}%
\bibitem [{\citenamefont {Jaber-Urquiza}\ \emph {et~al.}(2019)\citenamefont
  {Jaber-Urquiza}, \citenamefont {Piccinelli},\ and\ \citenamefont
  {S\'anchez}}]{Jaber-Urquiza:2018oex}%
  \BibitemOpen
  \bibfield  {author} {\bibinfo {author} {\bibfnamefont {J.}~\bibnamefont
  {Jaber-Urquiza}}, \bibinfo {author} {\bibfnamefont {G.}~\bibnamefont
  {Piccinelli}}, \ and\ \bibinfo {author} {\bibfnamefont {A.}~\bibnamefont
  {S\'anchez}},\ }\href {\doibase 10.1103/PhysRevD.99.056011} {\bibfield
  {journal} {\bibinfo  {journal} {Phys. Rev. D}\ }\textbf {\bibinfo {volume}
  {99}},\ \bibinfo {pages} {056011} (\bibinfo {year} {2019})},\ \Eprint
  {http://arxiv.org/abs/1810.05708} {arXiv:1810.05708 [hep-ph]} \BibitemShut
  {NoStop}%
\end{thebibliography}%

\end{document}